\documentclass[12pt,preprint]{aastex63}
\usepackage{amssymb}
\usepackage{amsmath}
\usepackage{epstopdf}
\usepackage{xcolor}
\usepackage{amsmath}
\usepackage{graphicx}
\usepackage{hyperref}
\usepackage{graphics,graphicx,natbib}
\usepackage{subfigure}
\usepackage{color}

\usepackage{booktabs} 

\def\lapp{\ifmmode\stackrel{<}{_{\sim}}\else$\stackrel{<}{_{\sim}}$\fi}
\def\gapp{\ifmmode\stackrel{>}{_{\sim}}\else$\stackrel{>}{_{\sim}}$\fi}

\newcommand{\degrees}{^{\circ}}

\begin{document}

\title{A nearby repeating fast radio burst in the direction of M81}

\newcommand{\mcgillphysics}{Department of Physics, McGill University, 3600 rue University, Montr\'eal, QC H3A 2T8, Canada}
\newcommand{\msi}{McGill Space Institute, McGill University, 3550 rue University, Montr\'eal, QC H3A 2A7, Canada}
\newcommand{\wvuphysics}{Department of Physics and Astronomy, West Virginia University, P.O. Box 6315, Morgantown, WV 26506, USA}
\newcommand{\wvugws}{Center for Gravitational Waves and Cosmology, West Virginia University, Chestnut Ridge Research Building, Morgantown, WV 26505, USA}
\newcommand{\uoftphysics}{Department of Physics, University of Toronto, 60 St. George Street, Toronto, ON M5S 1A7, Canada}
\newcommand{\cita}{Canadian Institute for Theoretical Astrophysics, 60 St. George Street, Toronto, ON M5S 3H8, Canada}
\newcommand{\dunlapinstitute}{Dunlap Institute for Astronomy \& Astrophysics, University of Toronto, 50 St. George Street, Toronto, ON M5S 3H4, Canada}
\newcommand{\dunlapdep}{David A. Dunlap Department of Astronomy \& Astrophysics, University of Toronto, 50 St. George Street, Toronto, ON M5S 3H4, Canada}

\newcommand{\mitkavli}{MIT Kavli Institute for Astrophysics and Space Research, Massachusetts Institute of Technology, 77 Massachusetts Ave, Cambridge, MA 02139, USA}
\newcommand{\mitphysics}{Department of Physics, Massachusetts Institute of Technology, 77 Massachusetts Ave, Cambridge, MA 02139, USA}
\newcommand{\ubc}{Dept. of Physics and Astronomy, 6224 Agricultural Road, Vancouver, BC V6T 1Z1 Canada}
\newcommand{\perimeter}{Perimeter Institute for Theoretical Physics, 31 Caroline Street N, Waterloo ON N2L 2Y5 Canada}
\newcommand{\tata}{Department of Astronomy and Astrophysics, Tata Institute of Fundamental Research, Mumbai, 400005, India}
\newcommand{\ncra}{National Centre for Radio Astrophysics, Post Bag 3, Ganeshkhind, Pune, 411007, India}
\newcommand{\drao}{Dominion Radio Astrophysical Observatory, Herzberg Astronomy \& Astrophysics Research Centre, National Research Council Canada, PO Box 248, Penticton, BC V2A 6J9, Canada}

\newcommand{\waterloo}{Department of Physics and Astronomy, University of Waterloo, Waterloo, ON N2L 3G1, Canada}


\author[0000-0002-3615-3514]{M.~Bhardwaj}
\affiliation{\mcgillphysics}
\affiliation{\msi}

\author[0000-0002-3382-9558]{ B.~M.~Gaensler}
\affiliation{\dunlapinstitute}
\affiliation{\dunlapdep}

\author[0000-0001-9345-0307]{V.~M.~Kaspi}
\affiliation{\mcgillphysics}
\affiliation{\msi}

\author[0000-0001-9345-0307]{T.~L.~Landecker}
\affiliation{\drao}

\author[0000-0001-7348-6900]{R.~Mckinven}
\affiliation{\dunlapinstitute}
\affiliation{\dunlapdep}

\author[0000-0002-2551-7554]{D.~Michilli}
\affiliation{\mcgillphysics}
\affiliation{\msi}

\author[0000-0002-4795-697X]{Z.~Pleunis}
\affiliation{\mcgillphysics}
\affiliation{\msi}


\author[0000-0003-2548-2926]{S.~P.~Tendulkar}
\affiliation{\tata}
\affiliation{\ncra}

\author[0000-0001-5908-3152]{B.~C.~Andersen}
\affiliation{\mcgillphysics}
\affiliation{\msi}

\author[0000-0001-8537-9299]{P.~J.~Boyle}
\affiliation{\mcgillphysics}
\affiliation{\msi}

\author[0000-0003-2047-5276]{T.~Cassanelli}
\affiliation{\dunlapinstitute}
\affiliation{\dunlapdep}

\author[0000-0002-3426-7606]{P.~Chawla}
\affiliation{\mcgillphysics}
\affiliation{\msi}

\author[0000-0001-6422-8125]{A.~Cook}
\affiliation{\dunlapinstitute}
\affiliation{\dunlapdep}

\author[0000-0001-7166-6422]{M.~Dobbs}
\affiliation{\mcgillphysics}
\affiliation{\msi}

\author[0000-0001-8384-5049]{E. Fonseca}
\affiliation{\mcgillphysics}
\affiliation{\msi}
\affiliation{\wvuphysics}
\affiliation{\wvugws}

\author[0000-0003-4810-7803]{J.~Kaczmarek}
\affiliation{\drao}

\author[0000-0002-4209-7408]{C. Leung}
\affiliation{\mitkavli}
\affiliation{\mitphysics}

\author[0000-0002-4279-6946]{K.~Masui}
\affiliation{\mitkavli}
\affiliation{\mitphysics}

\author[0000-0002-3777-7791]{M.~Münchmeyer}
\affiliation{\perimeter}

\author[0000-0002-3616-5160]{C.~Ng}
\affiliation{\dunlapinstitute}

\author{M.~Rafiei-Ravandi}
\affiliation{\perimeter}
\affiliation{\waterloo}

\author[0000-0002-7374-7119]{P.~Scholz}
\affiliation{\dunlapinstitute}

\author[0000-0002-6823-2073]{K.~Shin}
\affiliation{\mitkavli}
\affiliation{\mitphysics}

\author{K.~M.~Smith}
\affiliation{\perimeter}

\author[0000-0001-9784-8670]{I.~H.~Stairs}
\affiliation{\ubc}

\author[0000-0001-8278-1936]{A.~V.~Zwaniga}
\affiliation{\mcgillphysics}
\affiliation{\msi}

\correspondingauthor{Mohit Bhardwaj}
\email{mohit.bhardwaj@mail.mcgill.ca}


\begin{abstract}

We report on the discovery of FRB~20200120E, a repeating fast radio burst (FRB) with low dispersion measure (DM), detected by the Canadian Hydrogen Intensity Mapping Experiment (CHIME)/FRB 
project. The source DM of 87.82 pc cm$^{-3}$ is 
the lowest recorded from an FRB to date,
yet is 
significantly
higher than the maximum 
expected from the Milky Way interstellar medium in this direction 
($\sim$ 50 pc cm$^{-3}$). We have detected three bursts and one candidate burst from the source over the period 2020 January-November. The baseband voltage data for the event on 2020 January 20 enabled a sky localization of the source to within $\simeq$~14 sq. arcmin (90\% confidence).  
The FRB localization is close to 
M81, a spiral galaxy at a distance of 3.6 Mpc.
The FRB appears on the outskirts of M81 (projected offset $\sim$~20 kpc) but well inside its extended H$\textsc{i}$ and thick disks. We empirically estimate the probability of chance coincidence with M81 to be $< 10^{-2}$.
However, we cannot 
reject a Milky Way halo origin for the FRB.
%
Within the FRB localization region, we find several interesting cataloged M81 sources and a radio point source detected in the Very Large Array Sky Survey (VLASS). 
We searched for prompt X-ray counterparts in {\it Swift}/BAT and {\it{Fermi}}/GBM data, and for two of the FRB~20200120E bursts, we rule out coincident SGR 1806$-$20-like X-ray bursts.
Due to the proximity of FRB 20200120E, future follow-up for prompt multi-wavelength counterparts 
and sub-arcsecond localization 
could be constraining of proposed FRB models.
\end{abstract}


\section{Introduction}

Fast radio bursts (FRBs) are millisecond-duration, bright 
radio transients with unknown physical origins \citep{lbm+07,tsb+13}.
Although more than 500 bursts have been reported thus far (CHIME/FRB collaboration et al., submitted), 
only 13 published FRBs have been sufficiently well localized on the sky to allow host galaxies to be identified.  The host galaxies of the localized FRBs have redshifts ranging from 0.03 to 0.66, demonstrating the 
significant distances at which FRBs are located\footnote{\url{http://frbhosts.org/} (visited on 19/12/2020)}.
A small fraction of  FRBs have been observed to repeat \citep{spitler2016repeating,2019Natur.566..235C, andersen2019chime, fonseca2020nine,kumar19:askap}, ruling out cataclysmic models at least for these sources. Among the repeating FRBs, two have shown periodic repetition of as yet unknown origin \citep{Amiri:2020gno,Rajwade:2020uat,2021MNRAS.500..448C}.

With this diverse phenomenology, understanding the origins of FRBs is a topical problem in astronomy.
One promising method to unravel FRB physical origins is to study their hosts and local environments and compare them with those of the proposed progenitors \citep{2017ApJ...843...84N,2020ApJ...899L...6L}. However, FRBs, both repeating and apparently non-repeating, are found in a variety of types of host galaxies and it is not yet clear if the two populations are intrinsically different \citep{bhandari2020host, heintz2020host}. Therefore, more host localizations are crucial to understand the nature of FRB sources. 
    
The recent discovery of FRB-like radio bursts \citep{2020SGR,Bochenek2020} and coincident X-ray bursts \citep{mereghetti2020integral,li2020identification,ridnaia2020peculiar} from Galactic magnetar SGR 1935+2154 argue strongly that at least some FRB sources are magnetars. Contemporaneous X-ray emission from SGR 1935+2154 means FRBs may not be solely a radio phenomenon. 
However, to test different proposed FRB progenitor models, multi-wavelength follow-up programs are most constraining for nearby sources ($< 50$ Mpc) due to the sensitivity limitations of high energy telescopes \citep[e.g.][]{scholz2020simultaneous}. Therefore, local universe FRBs are excellent candidates to test and constrain different proposed FRB models \citep[for a comprehensive list, see][and  \url{frbtheorycat.org}]{platts2019living}, and to study in detail the local environments of the FRB sources \citep{marcote2020repeating,2020arXiv201211617M,2020arXiv201103257T}.

Here we report the discovery of FRB 20200120E, a repeating FRB detected by the 
CHIME/FRB project \citep[][CHIMEFRB18 hereafter]{2018ApJ...863...48C}. The FRB dispersion measure (DM) of 87.82 pc cm$^{-3}$ is the smallest reported for an FRB thus far.
In \S\ref{Sec:obs}, we report on FRB 20200120E and its
three bursts and one candidate burst.
In \S\ref{sec:host}, we present evidence against the source being associated with the Milky Way or its halo, though the halo origin cannot yet be rejected conclusively.
We then describe our search for a host and propose that M81, a nearby early-type grand-design spiral galaxy at 3.63 $\pm$ 0.34 Mpc \citep{freedman1994hubble,karachentsev2002m}, is the most likely FRB host.
We further show the FRB is unlikely to be beyond M81.
In \S\ref{Sec:discussion}, we 
discuss the implications of this study and conclude in \S\ref{Sec:conclude}.

\section{Observations \& Analysis}
\label{Sec:obs}

\begin{figure}[!htb]
\centering
     \includegraphics{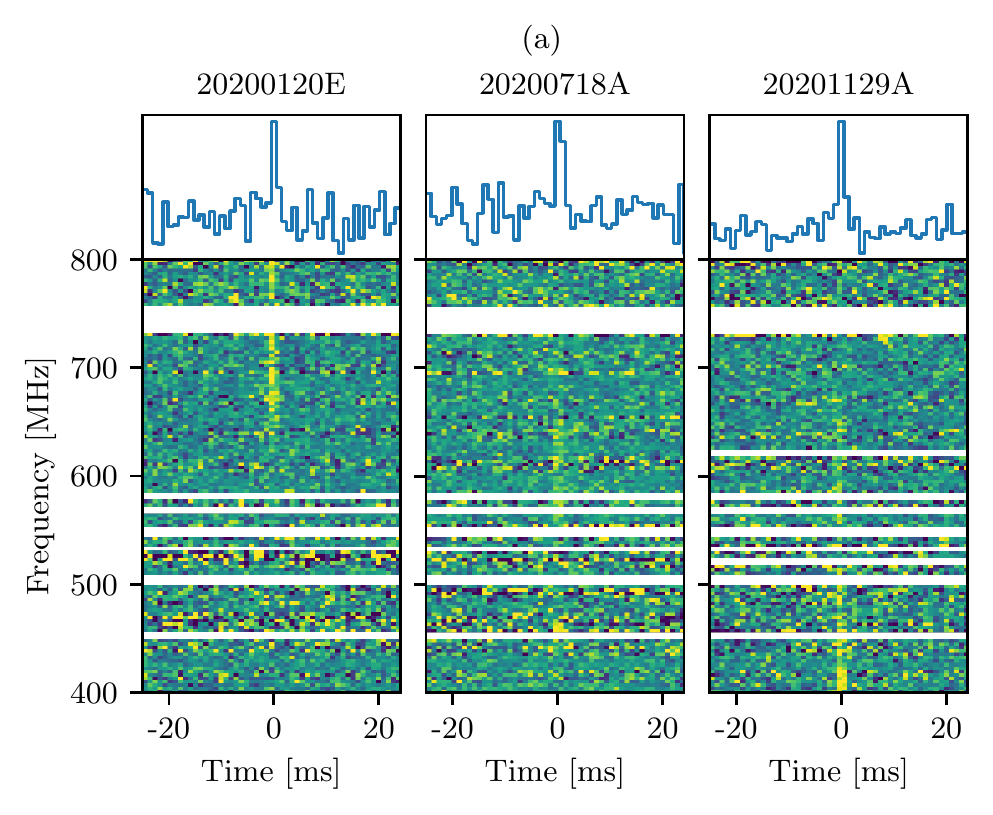}
     \includegraphics{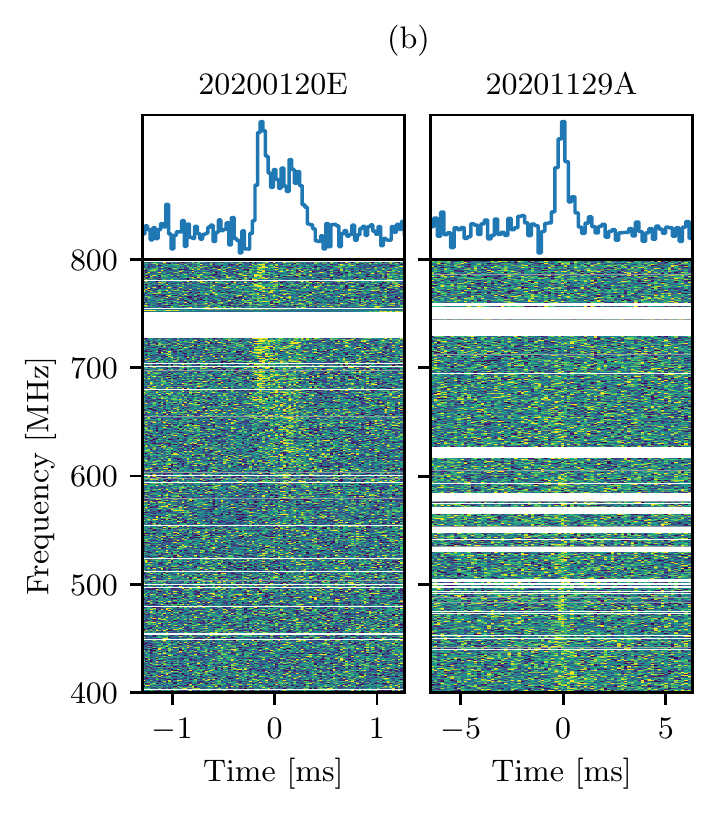}
\caption{Frequency versus time (``waterfall") plots of three dedispersed bursts detected from FRB 20200120E. \textbf{(a)} The dedispersed waterfall plots of the three bursts made using total-intensity data, with temporal and spectral resolutions of 0.983 ms, and 3.125 MHz, respectively. See Table \ref{ta:bursts} for the burst properties. \textbf{(b)} The waterfall plots of two FRB 20200120E bursts, FRBs 20200120E and 20201129A, made using baseband data, which here are binned to have temporal resolution 0.0256 and 0.16384 ms, respectively, and spectral resolution 0.391 MHz. 
The horizontal white lines in the waterfall plots are flagged channels that are either bad (due to radio frequency interference) or are missing due to computer nodes in the CHIME system being offline. Lastly, the fourth possible burst did not have intensity or baseband data recorded (see \S \ref{repeat_bursts}), and therefore, not shown here.
}
\label{Fig:waterfall_plots}
\end{figure}

\subsection{The Repeating Source FRB 20200120E}
\label{repeat_bursts}
FRB 20200120E was discovered by the Canadian Hydrogen Intensity Mapping Experiment (CHIME)/FRB project on 2020 January 20 in the real-time pipeline.  The CHIME/FRB instrument searches 1,024
formed sky beams for dispersed single pulses in the frequency range 400--800 MHz in total-intensity data, with time resolution 0.983 ms and 16k frequency channels.  
See CHIMEFRB18 for a detailed description of the CHIME/FRB detection system.
Offline analysis of the burst total-intensity data
using {\tt fitburst} \citep{2019Natur.566..235C,andersen2019chime,2019Natur.566..230C,fonseca2020nine}
measured a DM of 87.782~$\pm$~0.003~pc~cm$^{-3}$ (see Tables \ref{ta:bursts} \& \ref{Tab1}).

This event was also recorded by the CHIME/FRB baseband system,
which records full voltage data at the Nyquist sampling rate of 800 MHz for all
1,024 CHIME dual polarizarion antennas. 
Using the recorded baseband data and our offline baseband localization pipeline \citep{Daniele2020}, which has been
calibrated and tested using known positions of Galactic radio pulsars,
we have localized the FRB to a sky area of $\approx$ 14 arcmin$^{2}$ (90\% confidence region; see Table \ref{Tab1}).
The baseband localization is consistent with that inferred from the multi-beam intensity data detections, but is more precise as the full-array baseband data are used to estimate the localization region. 

We detected three bursts and one candidate burst from the FRB source. For the second burst on 2020 July 18, the CHIME/FRB baseband system was undergoing maintenance, so we saved only total-intensity data at the nominal CHIME/FRB search resolution.  
On 2020 November 29, we detected a third burst for which baseband data were recorded. Following a procedure identical to that used for the January 20 burst, we estimate the 90\% confidence localization region of the burst as: R.A. $\rm{= 09^{h}57^{m}42.^{s}1\pm18.^{s}9}$ and Dec = $68\degrees48\arcmin57\arcsec \pm 01\arcmin39\arcsec$. As the baseband localization region of the FRB from the 2020 January 20 burst is completely inside the latter localization region, we used the baseband localization region of the FRB in all our follow-up analyses. 
Later, while searching the CHIME/FRB event database, we found a fourth burst detected on 2020 February 6, having SNR 10.6, above the 10$\sigma$ threshold that the CHIME/FRB pipeline uses to identify candidate FRB events (see CHIMEFRB18). All the header data metrics, which are discussed in \cite{2018ApJ...863...48C}, suggest that the source is astrophysical. However, due to an unidentified issue in the FRB data recording system, the intensity data of the 2020 February 6 burst were not saved. Therefore, we cannot confirm its astrophysical nature. Hence, we excluded it from the analyses in this paper. In Table \ref{ta:bursts}, we report basic properties of this burst from the header data. 

Because the best-fit DMs and sky positions of the three bursts for which we have intensity and/or baseband data saved to disk are consistent, we conclude that the FRB is a repeater.
Therefore, we adopt the Transient Name Server (TNS)\footnote{\url{https://wis-tns.weizmann.ac.il/}} FRB naming convention, which gives the source the name of the first detected burst, FRB 20200120E. The TNS names of the bursts on 2020 July 18 and 2020 November 29 are FRBs 20200718A and 20201129A, respectively. As the intensity data of the FRB detected on 2020 February 6 were not saved, we did not request a TNS name for the burst. 


\begin{table}[t]
\begin{center}
\caption{Properties of the bursts from FRB 20200120E.$^a$ }
\hspace{-1.in}
\resizebox{1.1\textwidth}{!}{ 
\begin{tabular}{cccccccccccc} 
\hline
TNS Name & Day & MJD & Arrival Time$^b$& SNR  &  DM & Width & Scattering Time & Fluence$^c$ & Peak Flux  Density$^c$ & DM$_\mathrm{bb}^d$  & RM\\
 & (yymmdd) &     & (UTC @ 400 MHz) &  & (pc~cm$^{-3}$) & (ms) & (ms @ 600 MHz)& (Jy ms) & (Jy) & (pc~cm$^{-3}$) & (rad~m$^{-2}$) \\
\hline
 20200120E$^f$ & 200120 & 58868 & 09:57:35.984(2) & 22.9 & 87.782(3)&  0.16(5) & $<$0.23 & 2.25(12) & 1.8(9) & 87.789(9) & -29.8(5)$^g$  \\
$-$ & 200206$^e$ & 58885 & 08:50:45 & 10.6 & 88(1) & $-$ & $-$ & $-$ & $-$ & $-$ & $-$\\
20200718A & 200718 & 59048 & 22:12:31.882(1) & 14.0 &  87.864(5) & 0.24(6) & $<$0.17 & 2.0(7)  & 1.1(5) & $-$   & $-$ \\
20201129A$^f$ & 201129 & 59182 & 13:31:29.8583(6) & 19.3 & 87.812(4) & $<$0.1 & 0.22(3) &  2.4(1.4) & 1.7(1.2) & 87.71(5) & -26.8(3)$^g$  \\
\hline
\end{tabular}}
\label{ta:bursts}
\end{center}
$^a$ Uncertainties are reported at the $1\sigma$ confidence level (cl). Reported upper limits are those of the $2\sigma$ cl.\\
$^b$ All burst times of arrival are topocentric.\\
$^c$ Fluence and peak flux density measurements represent lower bounds as we assumed that the bursts were detected at the center of their detection beams.\\
$^d$ Optimized DM for the burst detected in the baseband data.\\
$^e$ Single beam event with sky position consistent with the FRB baseband localization stated in Table \ref{Tab1}. However, the intensity data were not saved, so a TNS name has not been assigned. The reported DM and timestamp of the FRB were from the header data of the event.\\
$^f$Burst parameters were estimated using the total-intensity data. \\
$^{g}$ Both RM measurements have an additional systematic uncertainty of 1.0 rad m$^{-2}$.\\
\end{table}

\begin{table}[ht]
\begin{center}
\caption{Major Observables of FRB 20200120E.}
\begin{tabular}{@{} *2l @{}}\toprule
\textbf{Parameter} & \textbf{Value}\\\midrule
R.A.(J2000)$^a$ & $\rm{09^{h}57^{m}56^{s}.7 \pm 34^{s}.6}$ \\ 
Dec. (J2000)$^a$ & $\rm{68\degrees49\arcmin32\arcsec \pm 01\arcmin24\arcsec}$ \\
$l,b$$^b$ & 142.$\degrees$19, +41.$\degrees$22 \\
DM$^c$ &  87.818~$\pm$~0.007  pc cm$^{-3}$\\
DM$_{\mathrm{MW, NE2001}}^d$ & 40 pc cm$^{-3}$ \\
DM$_{\mathrm{MW, YMW16}}^d$ & 35 pc cm$^{-3}$ \\
DM$_{\mathrm{MW, WIM}}^e$ & 10$-$40 pc cm$^{-3}$ \\
DM$_{\mathrm{MW, N_{H}}}^f$ & 14$-$28 pc cm$^{-3}$ \\
DM$_{\mathrm{MW, halo}}^g$ & 30 pc cm$^{-3}$ \\
Max. distance$^{h}$ & $\lesssim$ 135 Mpc \\
RM$^i$ & $\rm{-28.3\pm0.6\pm1.0~rad\; m^{-2}}$  \\\bottomrule 
 \hline
\end{tabular}

\label{Tab1}
\end{center}
$^a$ FRB position determined from baseband data saved for FRB 20200120E using the technique described by \cite{Daniele2020}. The quoted uncertainty is the 90\% cl.

$^b$ Galactic longitude and latitude for the baseband localization central coordinates.

$^c$ Weighted average DM of the three observed bursts (excluding the 2020 February 6 burst; see Table \ref{ta:bursts}).

$^d$ Maximum DM model prediction along this line-of-sight for the NE2001 \citep{cordes2002ne2001} and YMW16 \citep{yao2017new} Galactic electron density distribution models. 

$^e$ DM contribution of the Milky Way assuming that the ISM in the FRB sight-line is dominated by diffuse warm ionized medium (WIM); see \S\ref{galactic-m81}.

$^f$ DM contribution of the Milky Way using the $\rm{N_{H}-DM}$ relation from \cite{he2013correlation}; see \S\ref{galactic-m81}.

$^g$ Milky Way halo prediction from the \cite{dolag2015constraints} hydrodynamic simulation. The \cite{yamasaki2020galactic} model predicts a similar value $\sim~35$ pc cm$^{-3}$. Note that both these values are smaller than the prediction from \cite{prochaska2019low}, 50$-$80 pc cm$^{-3}$.

$^h$ Maximum luminosity distance (90\%~confidence upper limit) estimated using the Macquart relation \citep{macquart2020census}; see \S \ref{host-search}. 

$^i$  Weighted average RM of the FRBs 20200120E and 20201129A. 

\end{table}
\subsection{Burst properties}
\label{frb:osbservation}

Figure \ref{Fig:waterfall_plots}a shows dedispersed waterfall plots of the three named bursts, made using the estimate of each burst’s DM, shown in Table \ref{ta:bursts}, which is optimized to maximize the burst SNR in the total-intensity data.  We  used  the  calibration  methods  described by \cite{andersen2019chime} and \cite{fonseca2020nine} to  determine  fluences of the three bursts (see Table~\ref{ta:bursts}), assuming the baseband localization position. Lastly, we employed the same modeling  procedures  that are discussed by \cite{andersen2019chime} and \cite{fonseca2020nine} for  estimating  widths,  arrival  times and scattering timescales from the calibrated total-intensity dynamic spectra of the three TNS named bursts. In the total-intensity data, the three named bursts show only one component, and therefore, we fitted a single-component profile. 

We also show dedispersed waterfall plots for the two bursts for which we have baseband data saved to disk in Figure \ref{Fig:waterfall_plots}b.
We separately optimized DMs of the bursts, FRBs 20200120E and 20201129A, by aligning sub-structure with the \texttt{DM\_phase}\footnote{\url{https://github.com/danielemichilli/DM_phase}} module \citep{dmphase}. We estimated best-fit DMs of $87.780\pm0.009$ and $87.71\pm0.05$ pc~cm$^{-3}$, respectively. In the Figure we downsampled the data to have temporal and spectral resolution of 0.0256 and 0.16384 ms and 0.391 MHz, respectively. The dynamic spectrum of FRB 20200120E reveals downward drifting 
time-frequency sub-structures that 
are thus far exclusively observed in the dynamic spectra of repeating FRBs \citep{2019Natur.566..235C,andersen2019chime,2019ApJ...876L..23H,fonseca2020nine,2020MNRAS.497.3335D}. 
A detailed burst analysis of the two bursts using the baseband data is beyond the scope of this paper and will be discussed in future work. 

\subsection{FRB Rotation Measure}
\label{frb-rm}
\begin{figure*}
\centering     
\subfigure{\label{fig3:a}\includegraphics[width=0.49\textwidth]{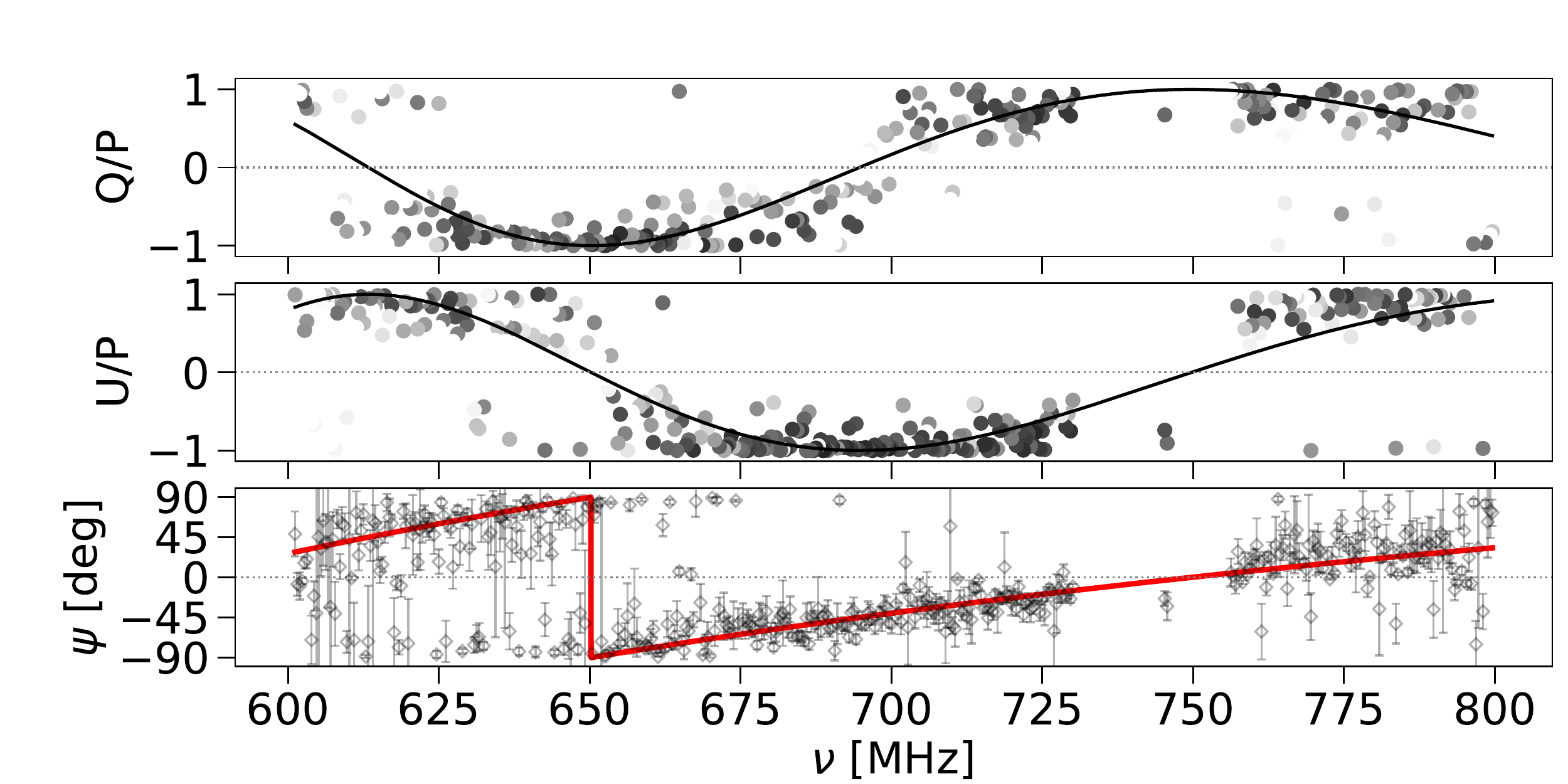}}
\subfigure{\label{fig3:b}\includegraphics[width=0.49\textwidth]{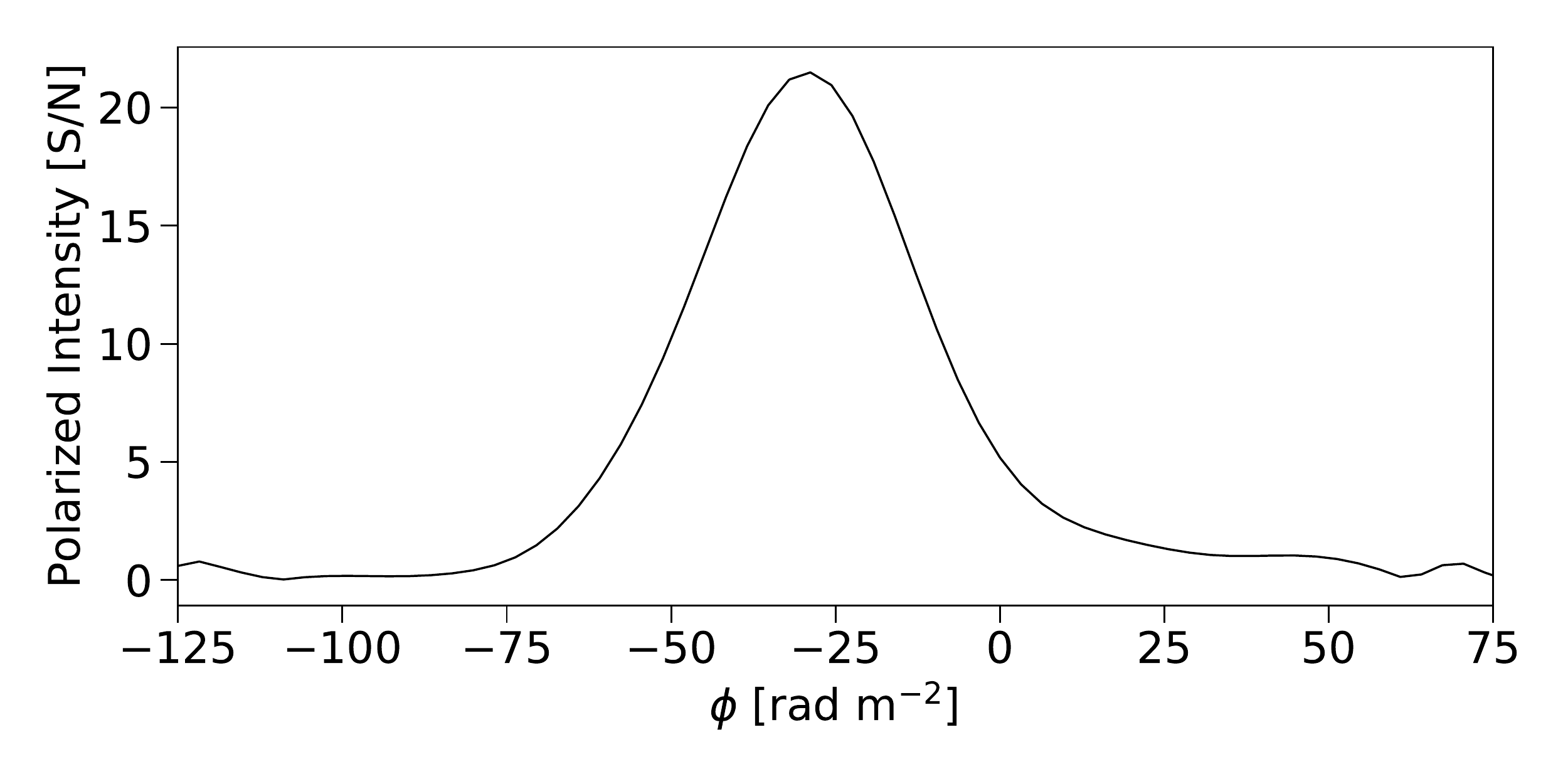}} \\

\subfigure{\label{fig3:c}\includegraphics[width=0.49\textwidth]{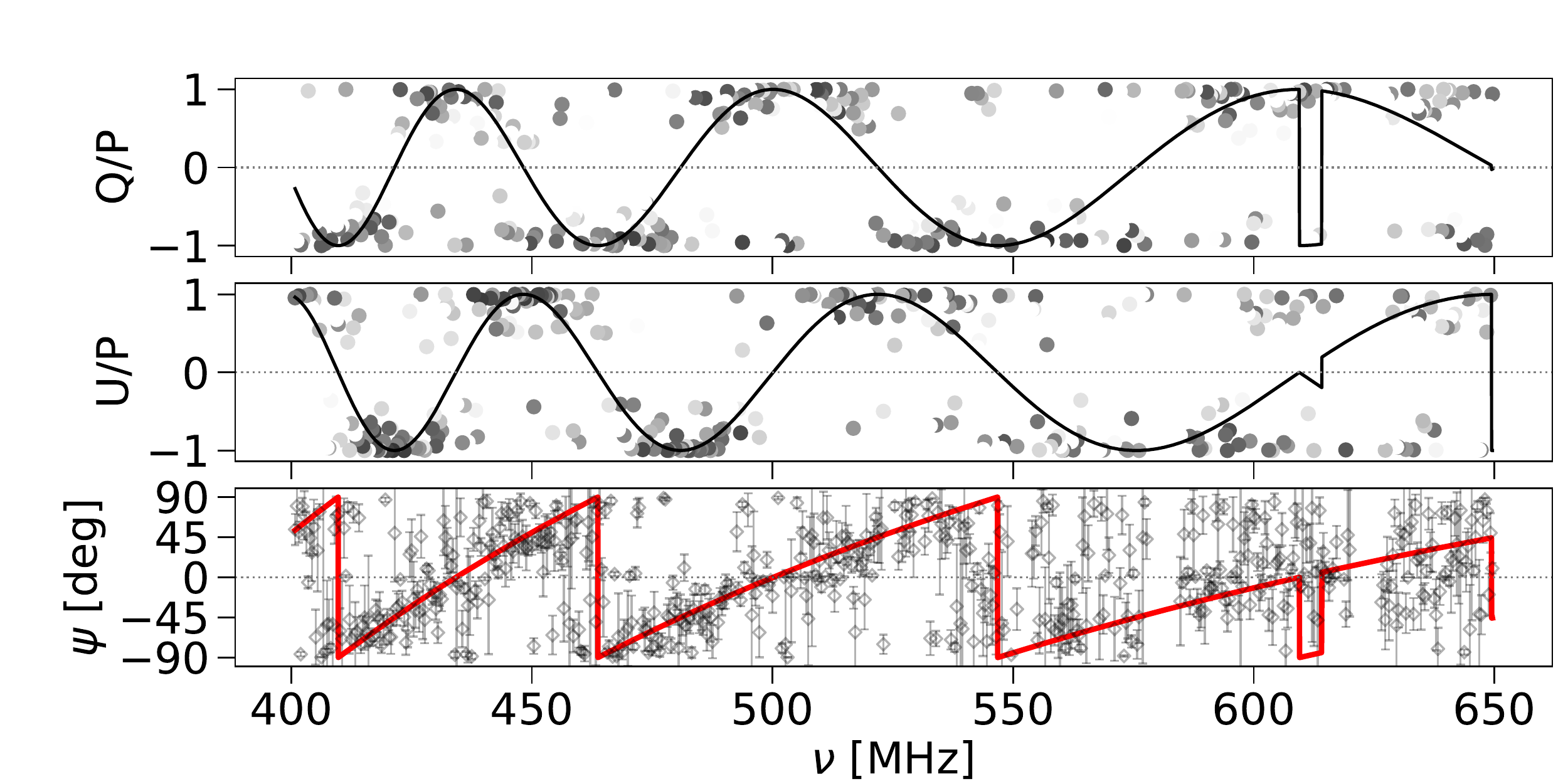}}
\subfigure{\label{fig3:d}\includegraphics[width=0.49\textwidth]{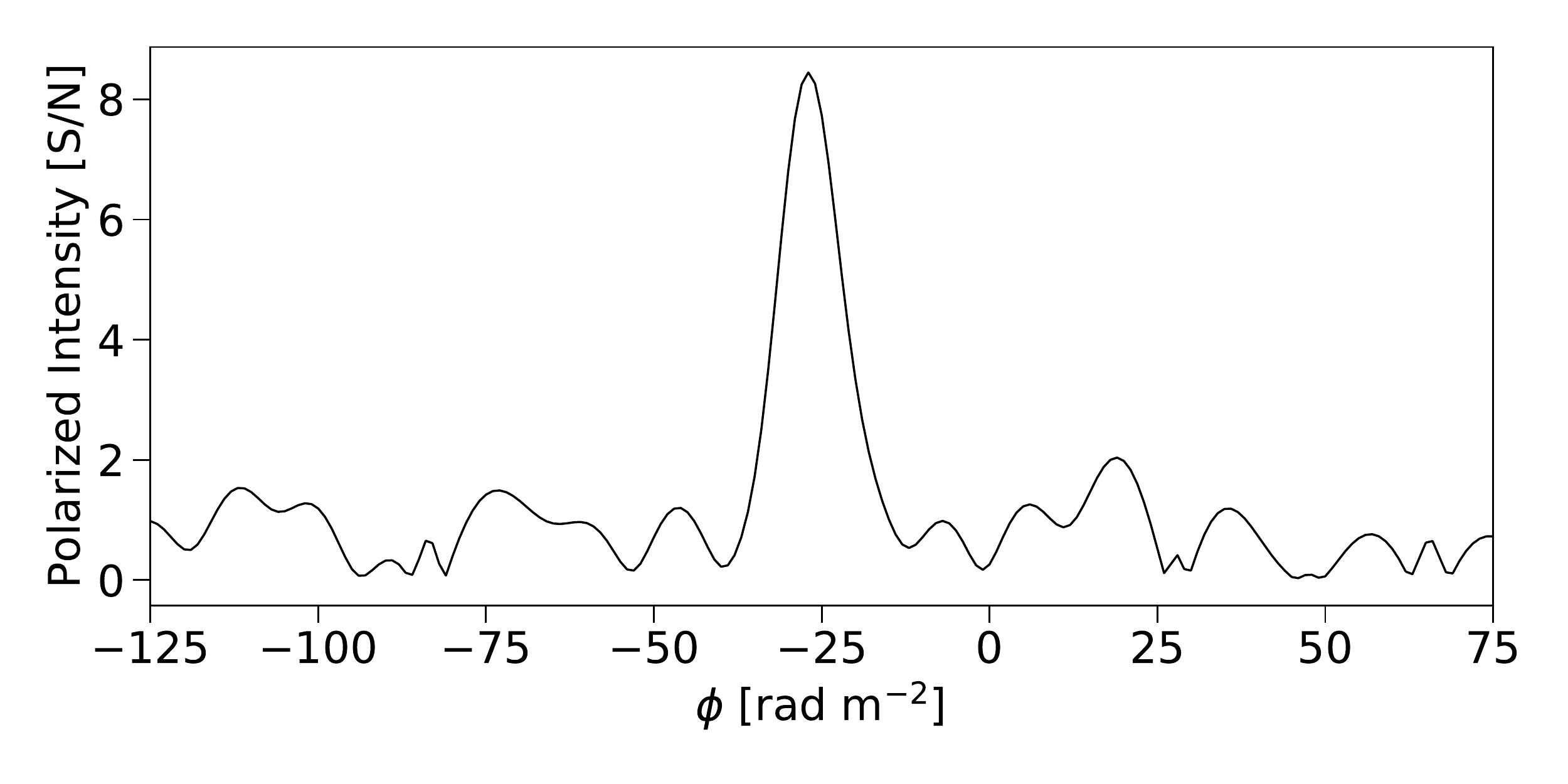}} 
\caption{Summary plots of the RM detection methods of QU-fitting and RM-synthesis applied to the cable delay corrected spectra of FRB 20200120E (top row) and FRB 20201129A (bottom row). Left panel: Stokes Q, U normalised by the total linear polarization ($\mathrm{P = \sqrt{Q^2+U^2}}$) and polarization angle, $\psi$, as a function of frequency with corresponding model fits. Frequency channels with significant polarized signal
are highlighted through a greyscale that saturates at higher S/N. Right panel: The results of RM-synthesis showing each event's FDF constrained near the peak.}
\label{fig:polarization}
\end{figure*}

Following a procedure similar to that outlined by \citet{fonseca2020nine}, a Faraday rotation measure (RM) for FRB 20200120E and FRB 20201129A were measured after applying RM-synthesis \citep{Burn1966,2005A&A...441.1217B} to the burst Stokes Q and U data and detecting a peak in the Faraday dispersion function (FDF). This initial measurement was refined by applying Stokes QU-fitting \citep{O'Sullivan2012}, modified to fit simultaneously for parameters characterizing the astrophysical signal as well as those corresponding to known systematics. In particular, a delay between the X and Y polarizations, arising from different path lengths through the electronics of the system (such as cable delay), produces mixing between Stokes U-V parameters (Mckinven et al., in prep.). A best-fit model is determined by a nested sampling implementation of QU-fitting that includes an additional parameter, $\tau$, characterizing the delay between the two polarizations. Leakage corrected spectra are shown in Figure~\ref{fig:polarization} along with model fits. 
Re-performing RM-synthesis on the cable-delay-corrected spectrum results in the FDFs shown in the right panel of Figure~\ref{fig:polarization}.
We 
estimate the RM of FRB 20200120E and FRB 20201129A to be $\rm{-29.8 \pm 0.5 \pm 1.0 \; rad\; m^{-2}}$ and $\rm{-26.8 \pm 0.3 \pm 1.0 \; rad\; m^{-2}}$. Quoted uncertainties correspond to the formal measurement and estimated systematic uncertainties, respectively. Ionospheric RM contributions have not been determined here but preliminary analysis indicates contributions of $\approx +(0.2-0.4) \mathrm{\; rad \, m^{-2}}$ \citep{ionFR2013}. 
The difference between these two measured RM values therefore is unlikely to be significant.
Moreover, both the bursts have nearly 100\% linear polarization fraction.
The Galactic foreground RM prediction in the FRB sight-line from \cite{oppermann2012improved} model is $\rm{ -11 \pm 8 \; rad\; m^{-2}}$. The low extragalactic RM and DM suggest that the FRB is unlikely to be located in a dense ionized region like a compact H\textsc{ ii}\ region or star-forming complex \citep{mitra2003effect,haverkorn2015magnetic,costa2018faraday,2018Natur.553..182M}, or in the Galactic center region of a host galaxy \citep{moss1996turbulence,krause2008magnetic}.   

\section{Determining the Distance and Host of FRB 2020120E}
\label{sec:host}
\subsection{Could FRB 20200120E be Galactic?}
\label{galactic-m81}
In this section, we discuss the possibility that FRB 20200120E 
is Galactic.
The maximum MW disk contribution to the DM along the FRB sight-line is 
40 pc cm$^{-3}$ from the NE2001 model \citep{cordes2002ne2001}, or 35 pc cm$^{-3}$
 from the YMW16 model \citep{yao2017new}.
The observed FRB DM is significantly larger than the DM predictions of 
either model even after taking into account an 20\% systematic uncertainty \citep{cordes2002ne2001,yao2017new}.
If the FRB is Galactic, an H\textsc{ ii} region could contribute to the DM-excess of the FRB as discussed by \cite{2018ApJ...869..181P}. We checked the \cite{anderson2014wise} H\textsc{ii}\ region catalog, which is claimed to be complete for Galactic H\textsc{ ii}\ regions other than large diffuse and young hypercompact H\textsc{ ii}\ regions, 
and found none, unsurprising given the high Galactic latitude of the source (b $= +41.\degrees22$; see Table~\ref{Tab1}). We also did not find any CO emission within the FRB localization region in the Planck all-sky CO map \citep{2014A&A...571A..13P} ruling out the presence of a young hypercompact H\textsc{ ii}\ region in a dense CO clumps of a molecular cloud complex \citep{2001ApJ...547..792D}. Finally, we searched the FRB field-of-view in the Wisconsin H$\alpha$ Mapper Northern Sky Survey \citep[WHAM;][]{haffner2003wisconsin} for the presence of any extended and diffuse H$\alpha$ excess clump but did not find any. Therefore, it seems unlikely that an H\textsc{ ii}\ region is responsible for the observed DM-excess.
 
 We consider two additional, independent maximum Galactic DM estimates. Using the observed Galactic hydrogen column density N$_{\rm{H}}$  in the FRB sight-line from the H\textsc{i} 4$\pi$ survey (HI4PI) \citep{2016A&A...594A.116H}, we find N$_{\rm{H}} =  5.8 \times 10^{20}$ cm$^{-2}$.  Using an empirically derived $\rm{N_{H}-DM}$ relation from \cite{he2013correlation}, we calculated the Milky Way contribution to the DM in the direction of the FRB to be $\sim$14$-$28 pc cm$^{-3}$, significantly smaller than the DM of the FRB. Note, however, that this N$_{\rm{H}}$--DM relation was estimated using radio pulsars, 
 which are generally located near to the Galactic plane.

We also independently estimated the Milky Way disk DM that, at high Galactic latitudes, is dominated by the warm ionized medium (WIM) extending to a vertical distance $\sim~$1$-$2 kpc above the Galactic plane \citep[e.g.,][]{2015HiA....16..574H}. From the Planck all-sky free-free emission map \citep{adam2016planck}, we find a total emission measure (EM) in the FRB sight-line of $\approx \rm{7.8~pc~cm^{-6}}$. To estimate the WIM-dominated MW disk DM, we used a range of free electron density ($N_{\rm{e}}$) values of the WIM from the literature \citep{1995ApJ...448..715R,reynolds2004warm,2008PASA...25..184G,2012A&A...541L..10V,2020ApJ...897..124O}, $N_{\rm{e}} \approx 0.2-0.9$~cm$^{-3}$, and computed DM $=$ EM$/N_{e} \sim~10-40~\rm{pc~cm^{-3}}$. Moreover, \cite{2007ASPC..365..250H} estimated the mean dispersion measure, $\overline{{\rm{DM}} \sin|b|}$, for high-EM sight-lines (EM~$\sin|b|>2$~pc~cm$^{-6}$) through the WIM to be $14.8\pm0.9$~pc~cm$^{-3}$. At this FRB's Galactic latitude, this relation gives a mean Galactic DM of $22.5 \pm 1.4$~pc~cm$^{-3}$.
From all these estimates, it seems highly unlikely the source is within the Milky Way disk. 

 The DM contribution of the MW halo, DM$_{\mathrm{MW, halo}}$, which consists primarily of hot ionized circumgalactic gas extending out to a Galactocentric radius of $\sim~$200 kpc, is poorly constrained \citep{keating2020exploring}. If we assume a halo DM contribution of 50$-$80 pc cm$^{-3}$ as proposed by \cite{prochaska2019low}, the FRB could
 be within the MW halo. 
However, two other MW halo DM models, those of \cite{yamasaki2020galactic} and \cite{dolag2015constraints}, predict DM$_{\mathrm{MW, halo}} \sim~$30 pc cm$^{-3}$, which support an extragalactic origin for the FRB.



If the FRB is a MW halo object, the DM-excess of $\sim~$50~pc~cm$^{-3}$ must be contributed by the hot coronal gas in the MW halo. Using different tracers of the hot gaseous medium \citep[$10^{5} - 10^{7}$ K;][]{tumlinson2017circumgalactic} that include emission and absorption lines of highly ionized species in the UV and X-ray \citep{2015ApJ...800...14M}
and constraints from the diffuse soft X-ray background \citep{2013ApJ...773...92H},
the electron density of the MW halo is estimated to be $\sim~10^{-3}$~cm$^{-3}$ close to the Milky Way disk \citep{2007ApJ...669..990B} and $\sim~10^{-4}$ cm$^{-3}$ at $\sim$~50 kpc \citep{2018ApJ...862....3B},
further reducing to $\sim~10^{-5}$ cm$^{-3}$ near the virial radius of the Milky Way \citep{Kaaret2020}.
Assuming a gas filling factor of unity and electron density of the coronal gas, $N_e \sim~10^{-3}-10^{-4}$~cm$^{-3}$, we estimate the distance to the FRB source to be $\sim$ 50--500 kpc. As the Milky Way virial radius is $\sim$~200 kpc \citep{2006MNRAS.369.1688D}, finding the FRB source at a distance $>$ 200 kpc implies that the source is extragalactic. Therefore, in further discussion, we use 50--200 kpc as a plausible distance range for a halo object. 

In this scenario, it is possible that the FRB source could be associated with either a Milky Way satellite galaxy \citep{2019IAUS..344..381K, 2019IAUS..344..377K} or globular cluster \citep{harris1996aj,2018yCat..36160012G,2019MNRAS.484.2832V}, but no such cataloged source exists within or near (within $\sim10\degrees$ radius circular sky area) to the FRB localization region \citep{2017MNRAS.466.1741C,2019ARA&A..57..375S}.

If the source is in the MW halo, the observed bursts 
could correspond to super-giant pulses (SGPs) from one of a young neutron star, millisecond pulsar, or rotating radio transient (RRAT).
We now consider each of these possibilities in turn.

It is difficult to explain the existence of a young neutron star in the halo, given that these objects are expected to form only near the Galactic Plane.
The fastest known runaway OB stars and young pulsars both have space velocities $\sim$~1000~km~s$^{-1}$ \citep{doi:10.1146/annurev-astro-082214-122230,2018ApJ...863...87D,chatterjee2005getting}, which would require $\sim~10^{8}$~yr to traverse $50-200~$kpc. This is  significantly longer than the typical lifetime of an OB star \citep[tens of Myrs;][]{10.1111/j.1468-4004.2012.53430.x} and inconsistent with the expectation of a young neutron star.
On the other hand, it has been suggested \citep[e.g.][]{giacomazzo2013formation} that young neutron stars can be formed by compact object mergers, which could occur in the halo. However, this formation mechanism has not been confirmed as actually occurring in nature.

Few millisecond pulsars (MSPs) are also known to emit SGPs \citep{2004IAUS..218..315J}. In principle, an isolated MSP can exist at a distance of $\sim$~50 kpc. 
However, Galactic isolated MSPs that are known to produce SGPs 
are rare; only two \citep[PSRs B1937+21 and B1821--24:][]{1996ApJ...457L..81C,2001ApJ...557L..93R} such sources have been seen to do so out of a sample of $\sim~450$ known MSPs \citep{manchester2005australia}. Moreover, 
assuming a SGP duration of $\sim~\mu$s, the maximum isotropic energy
observed from the brightest SGPs of PSR~B1937+21
is $\sim~10^{20}$ erg
\citep{2019MNRAS.483.4784M}. The isotropic burst energy of the FRB source, if located at a distance of $\sim$~50 kpc,
would be $\sim~10^{22}$ erg, 100 times brighter than the brightest giant pulses observed from PSR~B1937+21. Similar analysis using PSR~B1821--24 SGPs would give an even larger energy difference \citep{2013ApJ...778..106J}. Note that the energy difference would further increase if we assume the FRB source distance $>$ 50 kpc. 

Alternatively, the FRB source could potentially be explained as Crab-like
SGPs from a $\sim~10^8$ year old rotating radio transient (RRAT) in the halo.
However, it is unclear if an old RRAT can produce such pulses. For instance, the RRAT  with the largest isotropic energy estimated using the distance and mean flux density values from the RRAT catalogue\footnote{\url{http://astro.phys.wvu.edu/rratalog/} (visited on 19/12/2020)} is RRAT J1819$-$1458 with an energy of $\sim~10^{20}$ erg.
This is again two orders of magnitude below the energy needed to power FRB~20200120E at a distance of 50 kpc.

Finally, FRB 20200120E shows complex spectral and temporal downward-drifting sub-structures (see Figure \ref{Fig:waterfall_plots}b), 
which have previously been established as a characteristic spectro-temporal feature of repeating FRBs \citep{andersen2019chime,2019ApJ...876L..23H,fonseca2020nine,2020MNRAS.497.3335D}.
Such structures are not known to be seen in pulsar or RRAT spectra, although some Crab SGPs have shown similar complex structure \citep{2007ApJ...670..693H}.

Overall, if FRB 20200120E is a MW halo object, we conclude that it would be the most distant Galactic neutron star yet discovered
and would also need to be unusually energetic compared to known objects.

\begin{figure}[t]
\includegraphics[width=0.98\textwidth]{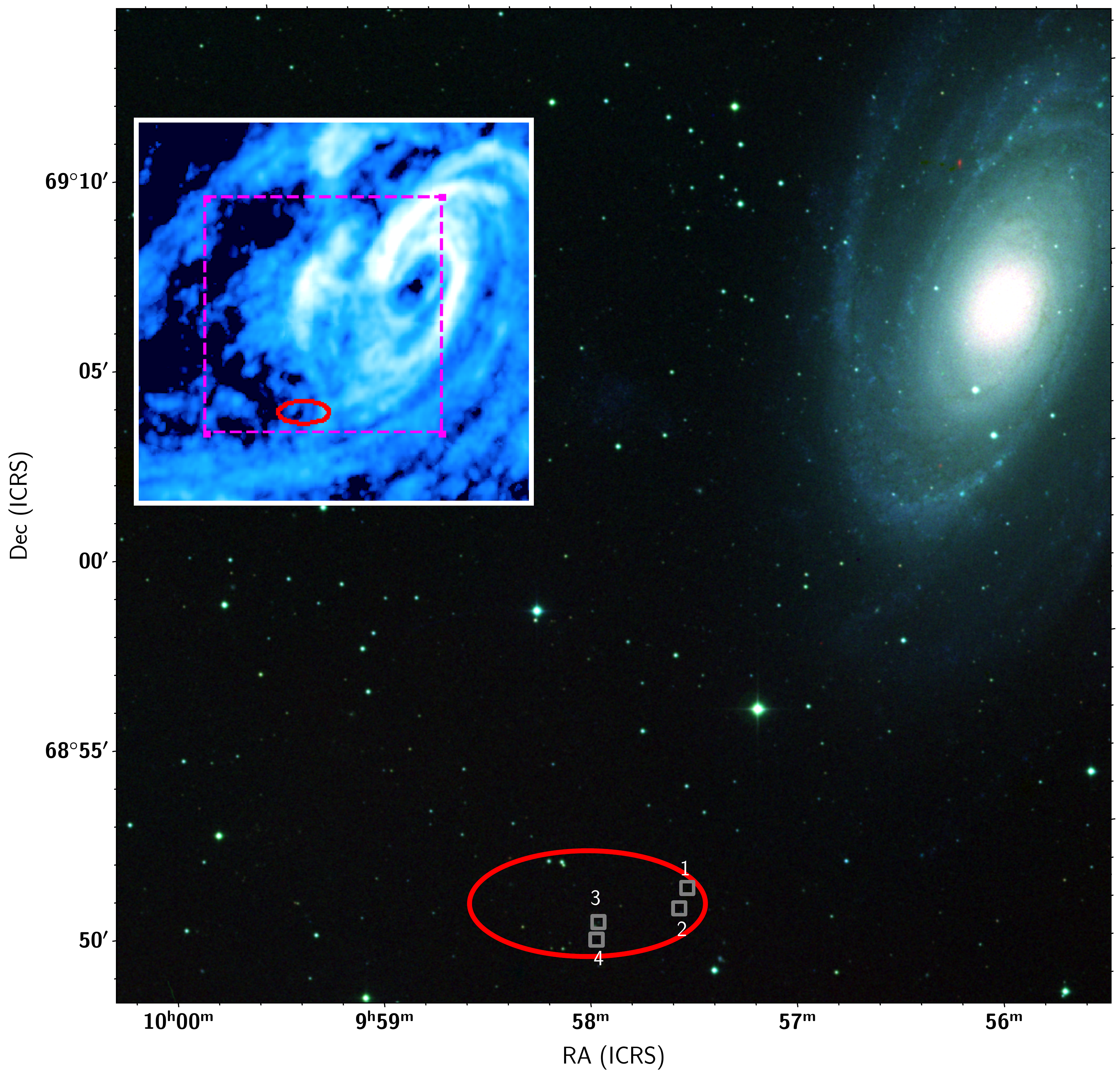}
\centering
\caption{The Digital Sky Survey (DSS) RGB image of the region around M81. The red ellipse represents the 90\% confidence localization region of FRB 20200120E. 
Source 1 is the cataloged M81 H\textsc{ ii} region, [PWK2012] 31 \citep{Patterson2012}, Source 2 is an X-ray source, [SPZ2011] 8 \citep{2011ApJ...735...26S}, Source 3 is an M81 globular cluster, [PR95] 30244, and Source 4 is the VLASS point radio source,  VLASS1QLCIR J095756.10+684833.3 \citep{2020RNAAS...4..175G}. All the sources are found in the outer disk of M81. The inset image is the 21-cm line view of the M81 circumgalactic medium \citep{chynoweth2008neutral}; the dashed magenta box is the DSS image field-of-view.} 
\label{Fig-M81FOV}
\end{figure}

\subsection{Host galaxy search}
\label{host-search}
In light of the challenges faced when associating FRB 20200120E with an object in the MW halo, we next consider whether it could be associated with an external galaxy.
If we assume DM$_{\mathrm{halo}}$  = 30 pc cm$^{-3}$ as predicted by the \cite{dolag2015constraints} and \cite{yamasaki2020galactic} models, the extragalactic DM of the FRB, DM$_{\rm{EG}}$, is 18 and $23 \; \rm{pc}\; \rm{cm}^{-3}$ for the NE2001 and YMY16 Milky Way DM models, respectively (see Table \ref{Tab1}).   For a negligible host DM contribution,
we estimate the maximum redshift of the FRB to be z$_{\mathrm{max}}$ $\approx$ 0.03 (90\% confidence upper limit), or maximum luminosity distance of 135 Mpc, using the Macquart DM$-$z relation \citep{mcquart-relation20}. 
Therefore, if the FRB is extragalactic, we expect a nearby host galaxy within its location error region. 

One possibility is that there is a faint dwarf
galaxy,
like that of FRB 121102, 
in the localization region of FRB 20200120E.
An FRB 121102-like star-forming dwarf galaxy 
\citep[M$_{\mathrm{r}} = -17$ 
AB mag;][]{tendulkar2017host}, 
if located at z$_{\mathrm{max}}$, would have r-band magnitude $\approx~$19 AB mag. Within the FRB 90\% confidence localization region, shown as a red ellipse in Figure \ref{Fig-M81FOV}, we identify one galaxy in the Dark Energy Spectroscopic Instrument (DESI) Legacy Imaging Survey photometric catalogue \citep{DESI-2019}, 2MASX J09575586+6848551 
with r-band magnitude = 15.35 AB mag.
However, it is known to be at z~= 0.19395(2) which is significantly further away than~z$_{\mathrm{max}}$ \citep{Huchra_2012}. Therefore, any dwarf host galaxy within the 90\% confidence region must be fainter than the FRB 121102 host. 

More interestingly, we find the FRB sight-line has a sky-offset from the M81 center, a nearby grand-design spiral galaxy, of $19.\arcmin6$. At the 3.6 Mpc distance of M81 \citep{karachentsev2002m}, this sky offset corresponds to a projected distance of 20$^{+3}_{-2}$ kpc\footnote{90\% c.l}  from the center of M81, 
well within the extended H\textsc{i} disk 
(see Figure \ref{Fig-M81FOV}) and thick disk of M81 \citep[$\approx 25$ kpc;][]{2005A&A...431..127T}. The FRB localization region is in the location where tidal interaction among M81 group members has resulted in the formation of star forming clumps.
Observations in the visible band give only marginal sign of these sources, but they are extensively studied in the radio and X-ray bands. These sources are discussed in \S \ref{M81gp_galaxies}. Moreover, between the M81 group and Milky Way, we did not find any cataloged field or Milky Way halo satellite galaxy. All these observations make M81 a plausible host galaxy for FRB 20200120E.

\subsection{Can the proximity to M81 be by chance?}
\label{pcc-section}
We now estimate the chance coincidence probability (P$_{\mathrm{cc}}$) of finding an M81-like bright galaxy close to the FRB localization region. 
We define P$_{\mathrm{cc}}$ $\rm{= A_{Gal}/A_{CHIME}}$, where $\rm{A_{Gal}}$ is the total angular sky area spanned by M81-like or brighter galaxies that are visible to CHIME, and $\rm{A_{CHIME}}$ is the total sky area visible to CHIME (sky area above Dec $\gtrsim -10\degrees$; $\approx$ 61\% of the total sky area).
To be conservative, we remove the Milky Way sightlines where the DM-excess of FRB 20200120E is less than the $\sim~$10\% systematic error on the maximum of the two different Milky Way DM model predictions \citep{cordes2002ne2001,yao2017new}, which we define as DM$\rm{_{ex}}$. This DM-excess constraint (hereafter C1) removes 10\% of the total sky visible to CHIME (mostly consisting of the Galactic plane), and we estimate $\rm{A_{CHIME}}$ = 20600 deg$^{2}$. Note that the CHIME sensitivity changes with declination, but this effect is likely insignificant in our case as all the nearby bright galaxies are within 30$\degrees$ of the zenith in the CHIME primary beam. Next, we use the catalog of the local volume galaxies\footnote{\url{https://www.sao.ru/lv/lvgdb/introduction.php} (visited on 19/12/2020)}, which is complete for M81-like bright galaxies, and find three galaxies other than M81 that have extinction corrected B-band magnitudes, m$_{B} \leq$ 7.5, m$_{\rm{B}}$ of M81\footnote{Though we have used B band magnitudes that are provided by the catalog, the results would not change if we use other optical band magnitudes.}: M31 (Andromeda, m$_{B}$ = 3.7 at 770 kpc), M33 (Triangulum, m$_{B}$ = 6.1 at 930 kpc), and IC 342 (m$_{B}$ = 7.2 at 3.28 Mpc). To estimate the total sky area of these galaxies, we use a circular region with an angular radius equivalent to a 20 kpc projected offset (of FRB 20200120E from M81) at their respective distances, which are $\approx$ 1.$\degrees$49, 1.$\degrees$23, 0.$\degrees$35,  and 0.$\degrees$31 for M31, M33, IC 342, and M81, respectively. Using these values we estimate $\rm{A_{Gal} \approx 12.4~deg^{2}}$, and hence, P$_{\mathrm{cc}}  = 6 \times 10^{-4}$.
\begin{figure*}[h]
\centering     
\includegraphics[width=0.9\textwidth]{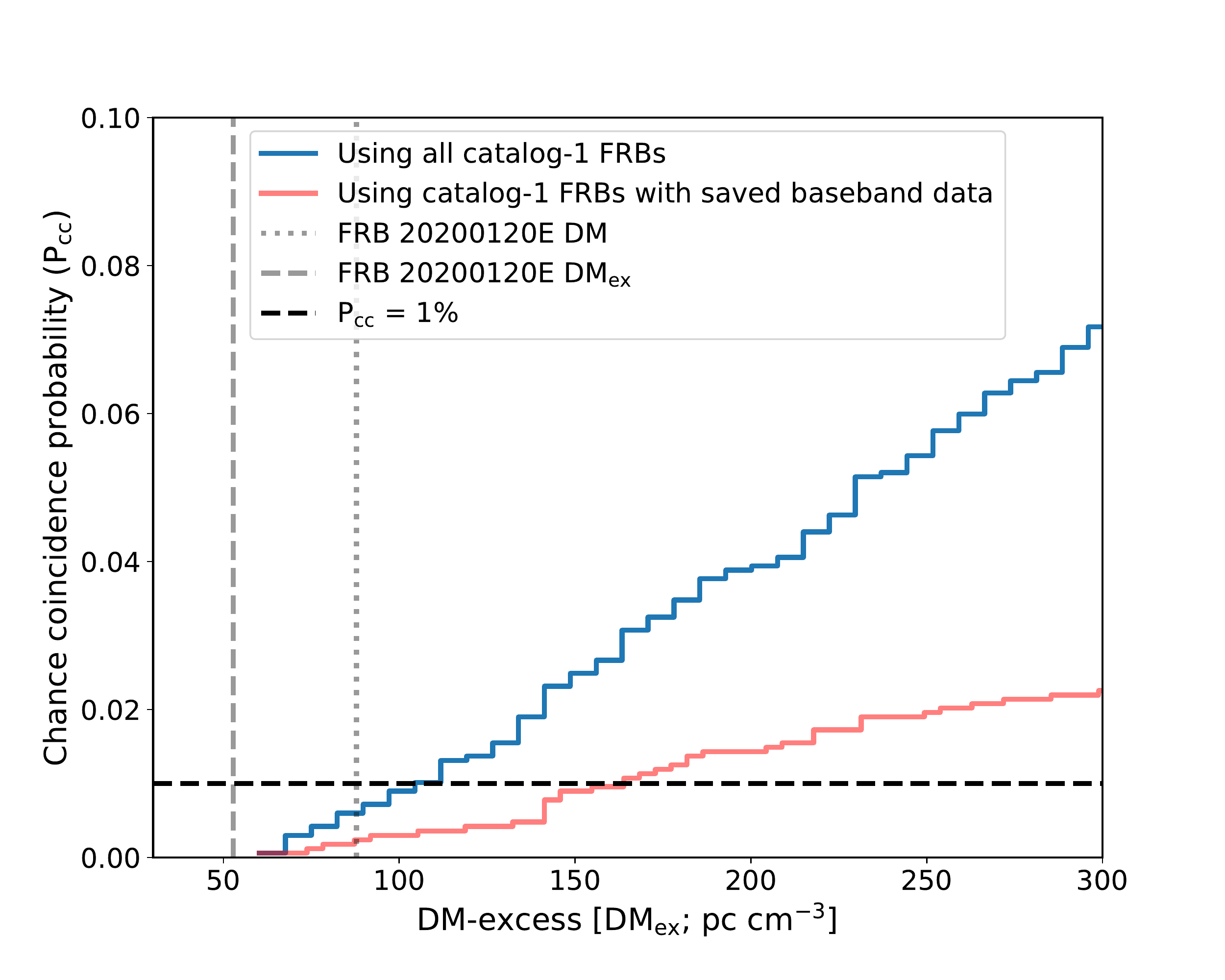}
caption{The chance coincidence probability of finding an M81-like galaxy as a function of the DM-excess (DM$\rm{_{ex}}$) of FRBs in the first CHIME/FRB catalog (submitted) that satisfy the DM-excess constraint, C1, as discussed in \S \ref{pcc-section}. This takes into account the look-elsewhere effect by incorporating the number of FRBs in the CHIME/FRB catalog with DM-excess $\leq$ DM$\rm{_{ex}}$ as a trial factor. FRB 20200120E has the lowest DM$\rm{_{ex}}$, which gives P$\mathrm{_{cc}}= 6\times 10^{-4}$. At the DM of FRB 20200120E = 87.82 pc cm$^{-3}$, P$\mathrm{_{cc}}$ = 0.7\%, which we consider as our conservative P$\mathrm{_{cc}}$ estimate. Moreover, the results would not change if we use minimum of the two different Milky Way DM model predictions \citep{cordes2002ne2001,yao2017new} to estimate DM$_{\rm{ex}}$ (instead of the maximum predicted values used in the DM-excess constraint C1). Lastly, we also showed the chance coincidence probability when only Catalog-1 FRBs with saved baseband data are used in correcting for the multiple testing problem.}
\label{fig:Pcc}
\end{figure*}

As the presence of M81 is inferred post-hoc, it is essential to correct the chance coincidence probability for the problem of multiple testing (also known as the look-elsewhere effect), that tends to increase the false-positive rate of a discovery \citep[Type I error;][]{maxwell2017designing}. To account for this, we use the Bonferroni correction procedure \citep{BENDER2001343,armstrong2014use}. We consider all FRBs in the first CHIME/FRB catalog (CHIME/FRB Collaboration et al., submitted) that satisfy the excess-DM constraint C1. The Bonferroni correction inflates the P$\mathrm{_{cc}}$ to $1 - (1 - 6\times10^{-4})^{\rm{N_{FRB,DM_{ex}}}}$, 
where N$\rm{_{FRB,DM_{ex}}}$ is the number of FRBs in the CHIME/FRB catalog with the DM-excess $\leq \rm{DM_{ex}}$. 
Figure \ref{fig:Pcc} shows the P$\mathrm{_{cc}}$ as a function of DM$_{\rm{ex}}$.
As FRB 20200120E has the lowest DM${\rm_{ex}}$ in our sample, N$\rm{_{FRB,47.8}}$ = 1, and therefore, P$_{\mathrm{cc}}$ = 6 $\times 10^{-4}$. To be conservative, we also count the number of CHIME FRBs with DM-excess $\leq$ 87.82 pc cm$^{-3}$, the DM of FRB 20200120E, and estimate N$\rm{_{FRB,87.82}} = 11$. This gives a value of P$_{\mathrm{cc}} =$ 0.007. 

There are several factors that make our P$_{\mathrm{cc}}$ estimate conservative. First, the Bonferroni correction becomes overly conservative as the number of events increase, and hence, undermines the significance of an unlikely observation \citep{perneger1998s,nakagawa2004farewell}. Second, we search prospective hosts for only those FRBs that have saved baseband data, which constitute a small fraction of the first CHIME/FRB catalog FRBs. Therefore, it could be argued that we should have used only them when accounting for the multiple testing problem 
\citep{streiner2011correction,maxwell2017designing}.
We also considered other nearby galaxies, except M81 satellite galaxies which are discussed in \S \ref{M81gp_galaxies}, that have projected angular offset less than or equal to that of M81 (19.\arcmin6) from the FRB, and for none is P$_{\mathrm{cc}} < 10$\%, even before correcting for the multiple testing problem. However, note that the formalism of chance coincidence probability favors brighter galaxies over the fainter ones. If FRBs preferentially originate in a specific galaxy type, then the P$_{\mathrm{cc}}$ is not a good proxy of true a host association probability. We need more host associations to test the latter possibility.

Bayesian hypothesis testing is another method proposed to avoid the problem of multiple testing \citep{scott2006exploration,gelman2012we}. However, its success in avoiding the problem of multiple testing strongly depends on the choice of priors \citep{rouder2014optional,de2020optional}. With better knowledge of the nature of FRB hosts, it will eventually be possible to use a Bayesian framework to estimate true host association probabilities.
 


\subsection{Could FRB 20200120E lie beyond M81?}
\label{m81-beyond?}
\begin{figure}[h]
\includegraphics[width=0.98\textwidth]{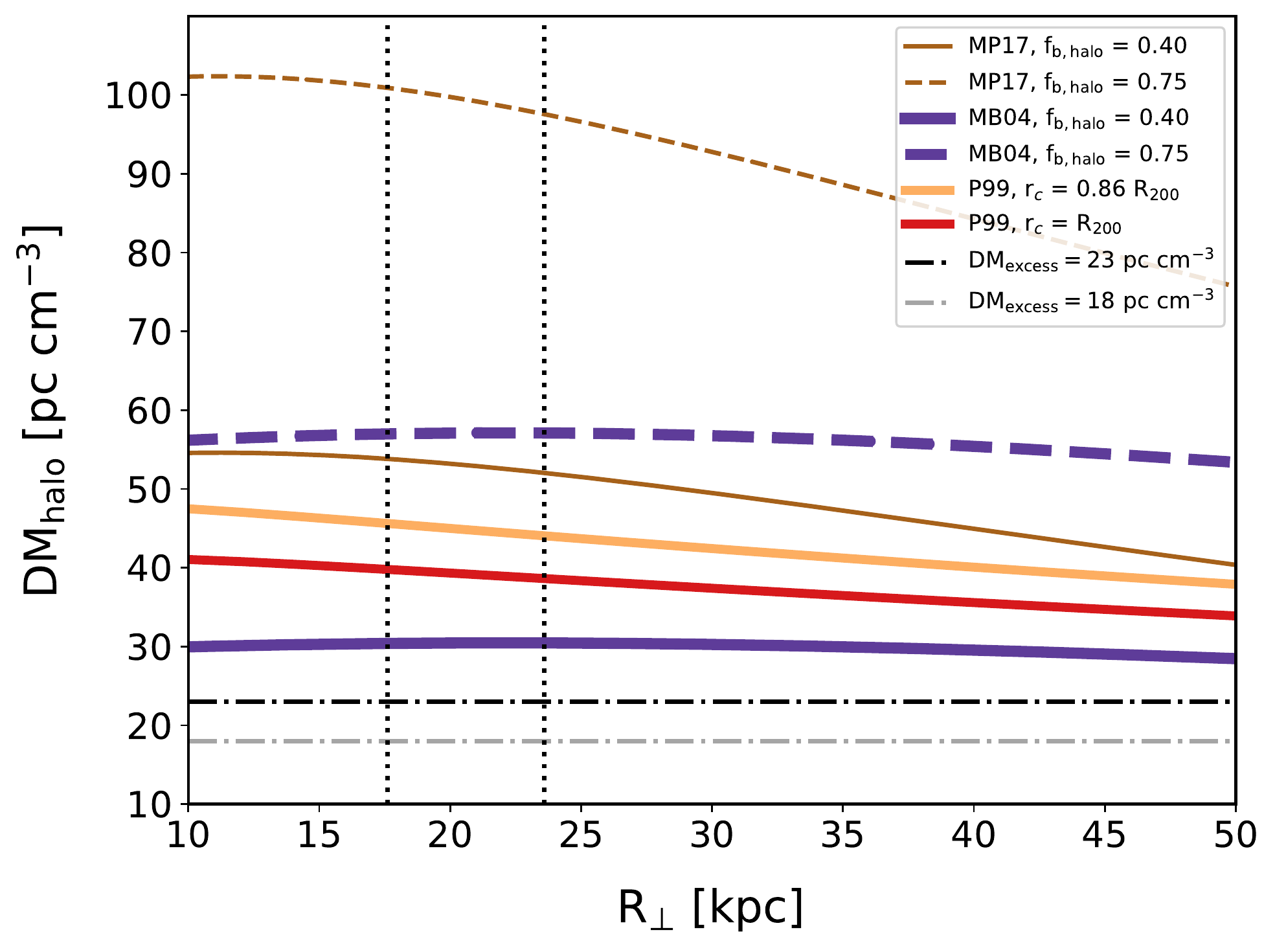}
\centering
\caption{M81 DM$_{\rm{halo}}$ as a function of impact parameter of the FRB assuming M81 is a foreground galaxy. 
We used two modified versions of the Navarro–Frenk–White (NFW) halo profile discussed by \cite{prochaska2019low}: MB04 ($\alpha$ = 2, $y_{0} =$ 4) and MP17 ($\alpha$ = 2, $y_{0} =2$); see \S\ref{m81-dm} for detailed description. The DM profiles are plotted for two M81 halo baryon fractions: (1) f$_{\mathrm{b,halo}} = 0.75$ (dash lines), a fiducial value that presumes that the halo has retained the mean cosmic baryons and that 25\% of these baryons are in the galaxy as stars, collapsed objects, and interstellar medium \citep[ISM;][]{fukugita1998cosmic};(2) f$_{\mathrm{b,halo}} = 0.4$ (solid lines), the minimum 
value that \cite{hafen2019origins} found in the FIRE simulation for a halo of mass $\sim~10^{12}$ $ \textup{M}_\odot$.
The region between the vertical dotted lines represents the range of the FRB projected distance from the center of M81 given the uncertainties in the FRB 90\% confidence localization region. 
As can be seen from the plot, even the most conservative scenario (MB04 profile with f$_{\mathrm{b,halo}} = 0.4$; solid purple line) predicts DM$_{\rm{halo}}$ greater than the DM-excess of FRB 20200120E, 18 and 23 pc cm$^{-3}$, using NE2001 and YMW16 model predictions, respectively. This suggests that it is unlikely that the FRB host is beyond M81. We also consider a non-NFW halo profile proposed by \cite{Pen_1999} (P99) with two core radius (r$\rm{_{c}}$) values from \cite{keating2020exploring}: $\rm{r_{c} = R_{200}~and~r_{c} = 0.86~R_{200}}$, where R$_{200}$ is the virial radius of the M81 halo; the large value of r$_{\rm{c}}$ allows more halo gas to be expelled from the M81 halo. Even in this scenario, we find the M81 $\rm{DM_{halo}}$ predicted by P99 model is significantly larger than the DM-excess of the FRB (see \S \ref{host-search}). Here we have included only the M81 halo contribution; adding contributions from other members of the M81 group would strengthen our conclusion. Note that if the FRB is associated with M81, the M81 DM$_{\rm{halo}}$ contribution would be likely smaller than the estimated values ($\lesssim$ 50\%).   
}
\label{Fig:Data2}
\end{figure}

\label{m81-dm}
Here we discuss the possibility of the FRB source being located beyond M81. In such a scenario, the CGM of M81 and its neighboring galaxies will also contribute to the FRB DM. 

M81 is a part of the nearby ``M81 group,'' which contains prominent galaxies like M82, NGC 2403, NGC 4236, and several dwarfs \citep{karachentsev2002m}. Some of its members are in the process of merging, which makes the M81 group CGM rich in metals and gas \citep{al2016ionized}. We estimate the DM contribution of the M81 halo assuming that the true host lies beyond M81. 

 The FRB 20200120E sight-line passes through the M81 halo with a very small impact parameter ($\sim~$20 kpc), which is much less than the M81 virial radius of $\sim~$210 kpc \citep{oehm2017constraints}. Therefore, we expect the M81 halo to contribute considerably to the FRB DM if the FRB is located beyond M81. To quantify this effect, 
 we assume a density profile and several major M81 halo parameters, such as halo mass and virial radius.  We first consider two halo gas profiles, proposed by \cite{maller2004multiphase} (hereafter MB04) and \cite{mathews2017circumgalactic} (hereafter MP17)\footnote{The \cite{dolag2015constraints} and \cite{yamasaki2020galactic} models are tailored to the Milky Way, and therefore, cannot be used here.}. Both profiles are modified versions of the Navarro-Frenk-White (NFW) density profile \citep{merritt2005universal}. For both profiles, we assume the following parametric form \citep{prochaska2019low}:
\begin{equation}
	\centering
	\rho = \frac{\rho_{\mathrm{0}}}{(\frac{\mathrm{r}}{\mathrm{R}_{\mathrm{s}}})^{1-\alpha}(\mathrm{y}_{0}+\frac{\mathrm{r}}{\mathrm{R}_{\mathrm{s}}})^{2+\alpha}}.
\label{Eq-density-profile}
\end{equation}
Here for the MB04 profile, we adopt $\alpha$ = 2, $y_{0} =$ 4, and for the MP17 profile,  $\alpha$ = 2, $y_{0} =2$. In Equation \eqref{Eq-density-profile}, $\rho_{0}$ is the central halo density, and $R_{\mathrm{s}} = R_{200}/c$, where $R_{200}$ is the scale radius that encloses a density of 200 times the critical density of the Universe, $\rho_{\mathrm{crit}} = 3H^{2}$/8$\pi G$, and $c$ is the concentration parameter defined as the ratio between the virial and scale radius of a halo. The values of these parameters for the M81 halo are $c = 10.29, R_{200}$ = 210 kpc, and $\rho_{0} = 7.21 \times 10^{-3} \; \textup{M}_\odot$ pc$^{-3}$, taken from \cite{oehm2017constraints}. As suggested by \cite{prochaska2019low}, we terminate the density profile at the M81 virial radius. Additionally, we need to assume a value for the fraction of baryons that is retained inside the virial radius of the M81 halo, f$_{\mathrm{b,halo}}$. For this, we consider two values: 
(1) f$_{\mathrm{b,halo}} = 0.75$ that assumes $\approx$ 25\% of the baryons exist in the galaxy as the interstellar medium (ISM), stars, and compact remnants \citep{1998ApJ...503..518F}, and (2) f$_{\mathrm{b,halo}}$ = 0.40 which is a lower limit that \cite{hafen2019origins} found for $\sim~10^{12}$M$_{\odot}$ mass halos in the Feedback in Realistic Environment (FIRE) simulation. The FIRE simulation is well suited to study the CGM of simulated galaxies with high resolution and also takes into account the effect of gas inflow and outflow along with other factors that play important roles in the evolution of galaxies \citep{10.1093/mnras/stu1738,10.1093/mnras/sty1690}. 
To convert the dark matter density to that of baryons, $\rm{\rho_{b}}$, we use the cosmic baryon fraction = $ \Omega_{\mathrm{b}}/\Omega_{\mathrm{\mathrm{m}}}$ = 0.158 \citep{ade2016planck}. Finally, we estimate the free-electron density using the relation: $\rm{n_{e} = f_{e}(\rho_{b}/m_{p}})$\footnote{We presume a flat $\Lambda$CDM \citep{2016A&A...594A..13P} model with the matter density, $\rm{\Omega_{m} = 0.308,~baryonic~matter~density, ~\Omega_{b} = 0.0486},~dark~energy~density$, $\Omega_{\Lambda} = 0.691$, and Hubble constant, H$_{0}$ = 100 h $\rm{km~s^{-1} Mpc^{-1}}$ with h = 0.6774.}, where $\rm{f_{e}}$ is the number ratio between free electrons and baryons in the halo \citep[$\approx$ 7/8;][]{tumlinson2017circumgalactic}, and m$_{\rm{p}}$ is the proton mass. Using these parameters, we estimate the M81 DM$_{\rm{halo}}$ using the following equation:

\begin{equation}
    \rm{DM_{\rm{halo}} = 2~\int_{0}^{\sqrt{r_{max}^{2} - R_{\perp}^{2}}} n_{e} dl.}
    \label{Eq-halo-DM}
\end{equation}

\noindent In Equation \eqref{Eq-halo-DM}, $\rm{r_{max}}$ is the maximum radius of integration through the M81 halo (R$_{200}$ in our analysis) and R$_{\perp}$ is the FRB impact parameter. 


Apart from the modified NFW-halo density profiles, we also consider the entropy-floor singular isothermal sphere model of \cite{Pen_1999}. The model invokes two phases of halo gas where in the inner region the gas is heated to constant entropy, and at radius (r) $\geq$ core radius (r$_{c}$), the gas traces the halo mass isothermally. The gas density profile is defined as :
\begin{equation}
    \rho(\rm{r}) = \Big(\frac{f_{g}v_{circ}^{2}}{4 \pi G}\Big)
    \left\{
	\begin{array}{ll}
		\rm{\frac{1}{r^{2}}  \& \mbox{if } r > r_{c},} \\
		\rm{\frac{1}{r^{2}_{c}}(1+\frac{12}{25}ln(\frac{r_{c}}{r}))^{1.5} \& \mbox{if }  r \leq r_{c}}
	\end{array}
    \right.
\end{equation}
Here $\rm{f_{g} = 0.06h^{-1.5}~and~v_{circ} = (10G H_{0} M_{h})^{1/3} \approx 154~km~s^{-1}}$ are the gas fraction and circular velocity of the M81 halo of mass M$_{\rm{h}} = 1.3\times 10^{12}$ M\textsubscript{\(\odot\)} \citep{oehm2017constraints} taken from \cite{Pen_1999} and \cite{mo1998formation}, respectively. 
We use two values of r$_{c}$ in our analysis as suggested by \cite{keating2020exploring}: $\rm{r_{c} = R_{200}~and~r_{c} = 0.86~R_{200}}$. To estimate the M81 halo DM, we use the procedure as discussed above.

Figure \ref{Fig:Data2} shows our estimates of the M81 DM$_{\mathrm{halo}}$ as a function of R$_{\perp}$. At the FRB R$_{\perp} \sim~$20 kpc, the minimum DM halo value (estimated using the MB04 profile and f$_{\mathrm{b,halo}}$ = 0.4) is larger than the FRB DM-excess of 18$-$23 pc cm$^{-3}$, see Table \ref{Tab1}).  This argues against the FRB being beyond or even on the far side of the halo. To check if our most constraining M81 DM$_{\rm{halo}}$ estimate is conservative, we estimated the Milky Way DM$_{\rm{halo}}$ using the MB04 profile, f$_{\mathrm{b,halo}}$ = 0.4, and other Milky Way halo parameters from \cite{prochaska2019probing}, and found DM$_{\rm{halo}} \approx$ 13 pc cm$^{-3}$.  This is smaller than the \cite{dolag2015constraints} and \cite{yamasaki2020galactic} predictions in the FRB sight-line, $\sim~$30 pc cm$^{-3}$. This demonstrates that our choice of halo density model does not bias the M81 DM$_{\rm{halo}}$ analysis. 

Note that in our analysis, we consider only the M81 halo contribution. However, other members of the M81 group would also contribute to the FRB DM. For instance, we can calculate DM$_{\mathrm{halo}}$ of the M81 group by only considering the two nearest massive M81 group members other than M81, NGC 3077 (R$_{\perp}$ $\sim~$36 kpc) and M82 (R$_{\perp} \sim~$59 kpc), with their respective halo parameters from \cite{oehm2017constraints}. The total DM$_{\mathrm{halo}}$ of the M81 group we estimate using the MB04 density profile with f$_{\mathrm{b}}$ = 0.4, the most conservative case in our analysis, is $\approx$ 75 pc cm$^{-3} >$ DM$_{\mathrm{EG}}$ = 53$-$48 pc cm$^{-3}$ (for a negligible Milky Way halo contribution).  This further strengthens the conclusion that the FRB is unlikely to be beyond the M81 group. Should the FRB source turn out to be behind M81, it would suggest that the M81 group has lost most of its halo baryons. 

As shown in Figure \ref{Fig-M81FOV}, the FRB localization region contains a significant amount of H$\textsc{i}$ gas. \cite{2005A&A...431..127T} argued that the M81 thick disk extends to a galactocentric radius of 25 kpc, in which case the observed H$\textsc{i}$ flux should be considered as a part of the M81 interstellar medium and hence, should make an additional DM contribution. In order to estimate the contribution of the H$\textsc{i}$ disk (DM$_{\mathrm{H\textsc{i}}}$), we use the N$_{\rm{H}}-$DM relation from \cite{he2013correlation}. As discussed in \S \ref{galactic-m81}, this relation is derived using nearby Milky Way radio pulsars, and so it is unclear whether it is valid for the M81 H$\textsc{i}$ disk. Noting the similarity between the Milky Way and M81 projected H$\textsc{i}$ distribution \citep{westpfahl1999geometry}, we use this relation for a rough estimate of DM$_{\mathrm{H\text{i}}}$. In the Very Large Array (VLA) image of the M81 group created by \cite{2018ApJ...865...26D}, the mean integrated H$\textsc{i}$ flux intensity within the 90\% localization region is 0.21 Jy~beam$^{-1}$ $\times$ km~s$^{-1}$. The VLA D-configuration beam is modelled as an ellipse with major ($\theta_{a}$) and minor ($\theta_{b}$) axes $38.\arcsec01$ and $30.\arcsec91$, respectively, and position angle = 75$\degrees$.5 that we get from the header data of the 21-cm map fits file made by \cite{2018ApJ...865...26D}\footnote{The image data can be directly downloaded from \url{https://www.astron.nl/~blok/M81data/}.}. Assuming the H$\textsc{i}$ gas to be optically thin, we estimate the integrated hydrogen column density, $\rm{N_{H}} \approx 2 \times 10^{20}~{\rm cm}^2$.

Using this N$\rm{_{H}}$ value in the N$\rm{_{H}}-$DM relation, we estimate DM$_{\mathrm{H\textsc{i}}} ~\rm{\sim~5-10~pc~ cm^{-3}}$, over and above the contribution from the halo, further diminishing the probability that the FRB lies beyond the M81 group. 

We conclude that FRB 20200120E
is unlikely to lie beyond the M81 group. Therefore, if the FRB is  an extragalactic source, its host is most likely located in the M81 group.

\subsection{Interesting M81 Group Sources}
\label{M81gp_galaxies}
In \S \ref{m81-beyond?}, we have shown that the FRB source is unlikely to be located beyond M81. Moreover, if we consider the MP17 halo density profile, the FRB is most likely to be located at the near side or in front of the M81 halo. However, with the MB04 and P99 halo density profiles, it is possible for the FRB to be present in the M81 extended H$\textsc{i}$ disk. In any case, we searched for any cataloged M81 group satellite galaxy
and did not find any 
within the FRB localization region. Within the M81 halo of sky-radius $\sim~3.\degrees3$ \citep[projected angular offset corresponding to the M81 virial radius, 210 kpc;][]{oehm2017constraints}, we found 42 dwarf satellite galaxies from the literature \citep{van1966luminosity,1985A&AS...60..213K,1968CoBAO..39...61K,1982AN....303..189B,1985A&AS...60..213K,caldwell1998dwarf,froebrich2000search,boyce2001blind,karachentsev2004new,Chiboucas_2009,smercina2017d1005+,okamoto2019stellar}. 
 We found a young dwarf irregular galaxy, Holmberg IX, and a dwarf spheroidal galaxy, KDG 64, at offsets of 13\farcm9 and 14\farcm3, respectively, from the center of the FRB localization region. Note that the two galaxies have the offset greater than the 10 times their half light radius. However, using the M81 satellite number density, $\rho_{\rm{g}} \approx$ 1.2 deg$^{-2}$, and assuming a Poisson distribution within the M81 halo as found in the semi-analytic and N-body gas dynamics studies by \cite{Kravtsov2004}, the chance coincidence probability (P$_{\rm{cc}}$) of finding these two galaxies near the FRB localization region can be estimated using the relation from \cite{bloom2002observed}, P$_{\rm{cc}}$ = 1 - e$^{\rm{-\rho_{g}A}}$, where A is the circular sky region of radius equals to the angular offset of the galaxies. We find 
P$_{\rm{cc}} > 10$\% for the galaxies.

Within or close to the FRB localization region, we found an M81 globular cluster, [PR95] 30244 \citep[Source 3 in Figure \ref{Fig-M81FOV};][]{2010AJ....139.2620N}. Though unremarkable, its presence in the FRB localization region is still noteworthy as globular clusters in the Milky Way are known to host several exotic systems, like millisecond pulsars, blue stragglers and low-mass X-ray binaries (LMXBs) \citep{2006ARA&A..44..193B}. Moreover, as shown in Figure \ref{Fig-M81FOV}, we found an H\textsc{ ii} region, [PWK2012] 31 (Source 1), and an X-ray source, [SPZ2011] 8 (Source 2), within the FRB localization region. 
The presence of an HII region confirms that in situ star formation is actively taking place within the extended H$\textsc{i}$ M81 disk. Additionally, \cite{2009MNRAS.399..737M} found a number of young stellar systems with ages of only a few tens of Myr in the M81 extended H$\textsc{i}$ disk, which were formed during the tidal interaction of M81 with surrounding companion galaxies, most prominently NGC 3077 and M82. Therefore, it seems possible that a young neutron star -- a possible FRB counterpart -- associated with M81 is present within
the FRB localization region. Lastly, \cite{2011ApJ...735...26S} estimated the counts in different X-ray sub-bands in the range 0.5$-$8 KeV. Using the X-ray color-based source classification first proposed by \cite{2003ApJ...595..719P}, if the X-ray source [SPZ2011] 8 is indeed an M81 source, we find that it can be either a high mass X-ray binary (HMXB), LMXB, or thermal supernova remnant. A future more precise FRB localization will tell us if the FRB is actually associated with any of these M81 sources. 

\subsection{Search for a persistent radio source}

We also searched for a persistent radio source, like the one seen coincident with FRB 121102 by \cite{chatterjee2017direct}, within the 90\% confidence localization region of the FRB. We searched archival data of the following surveys: the NRAO VLA Sky Survey  \citep[NVSS;][]{condon1998nrao}, the VLA Sky Survey  \citep[VLASS;][]{lacy2016vla}, the Westerbork Northern Sky Survey \citep[WENSS;][]{rengelink1997westerbork}, and the Tata Institute of Fundamental Research Giant Metrewave Radio Telescope Sky Survey \citep[TGSS;][]{intema2017gmrt}. From the search, we identified only one point source in the VLASS image
(2$-$4 GHz frequency band) above the 5$\sigma$ noise threshold ($\approx$ 0.6 mJy~beam$^{-1}$). In the Canadian Initiative for Radio Astronomy Data Analysis VLASS Epoch 1 Quick Look (CIRADA VLASS QL) Catalog \citep{2020RNAAS...4..175G}, this source is cataloged as VLASS1QLCIR J095756.10+684833.3 with 
R.A = $09^{\rm{h}}57^{\rm{m}}56^{\rm{s}}$ and Dec.= $+68^{\rm{d}}48^{\rm{m}}33^{\rm{s}}$ (Source 4 in Figure \ref{Fig-M81FOV}).
The estimated integrated and peak flux density of the source 
are 1.27 $\pm$ 0.19 mJy and 1.35 $\pm$ 0.11 mJy~beam$^{-1}$, respectively. Using the 5-$\sigma$ upper-limit from the NVSS image (1.4 GHz) of the FRB field-of-view of 2.15 mJy~beam$^{-1}$ and assuming a power-law dependence of the source flux density i.e., S$_{\nu} \propto \mathrm{S}^{\alpha}$, we estimated a lower limit $\alpha > -0.61$. Note that the VLASS calibrated images are known to have a systematic that underestimates the flux density of identified sources by $\sim~$10\% \citep{2020RNAAS...4..175G}. Incorporating this in our spectral index estimate would increase the derived lower limit.
Active galactic nuclei (AGN) are known to show variable light curves at all frequencies.  Given a $\approx$ 20-yr gap between the NVSS and VLASS observations, the derived spectral index upper limit is not constraining if the radio source turns out to be a background AGN.

We then compared the VLASS source 
with the one that was found spatially associated with FRB 121102 \citep{chatterjee2017direct}. At 3.6 Mpc, the isotropic luminosity 
of the radio source at 3 GHz, $\sim~10^{35} \; \rm{erg~s}^{-1}$, would be $\sim~10^{4}$ times smaller than that of the FRB 121102 radio source. The isotropic luminosity would be $\sim~10^{8}-10^{7}$ times smaller if the radio source is at distance 50$-$200 kpc. We can rule out a canonical stellar-mass X-ray binary, as such a source always has radio luminosity smaller than 10$^{33}$ erg~s$^{-1}$ at $\sim~$1 GHz even when flaring \citep{gallo2018hard,2020ApJ...888...36R}. 
A pulsar wind nebula (PWN) can give rise to a flat spectrum radio source \citep[$\alpha > - 0.4$;][]{2017SSRv..207..175R,2017hsn..book.2159S}. We searched for a possible optical counterpart of the radio source in the CFHT/MegaCAM image from \cite{2009AJ....137.3009C} and found none. The 5-$\sigma$ of r-band limit = 25.65 AB mag corresponds to an r-band flux of $1.5 \times 10^{36}$ erg~s$^{-1}$ at 3.6 Mpc. 
If the radio source is a Crab-like PWN with radio-to-optical luminosity ratio $\leq 10^{-3}$ \citep{volpi2008non}, it should have been detected in the CFHT/MegaCAM r-band image.
Moreover, \cite{2011ApJ...735...26S} observed M81, including the FRB field-of-view, using the {\it Chandra X-ray Observatory} and identified 276 X-ray point sources with sensitivity $\sim~10^{37}$ erg~s$^{-1}$. However, there is no X-ray point source spatially coincident with the radio source. If the radio source is an M81 supernova remnant or PWN, its X-ray/radio flux ratio is likely $< 10^{-2}$; otherwise, it is most likely a background AGN. Note that these constraints are also valid if the FRB source is located at 50$-$200 kpc. The persistent radio emission from X-ray binaries are orders of magnitude lower than their X-ray luminosities \citep{2019ApJ...871...26K}, therefore, a Galactic X-ray binary as the counterpart of the persistent radio source is also disfavoured from the constraints discussed above. Future follow-up observations of the persistent radio source and a milliarcsecond localization of the FRB may one day tell us if the radio source is associated with the FRB.   
%


\section{Discussion}
\label{Sec:discussion}
 
We have shown that the sky location of the low-DM repeating FRB 20200120E  appears superimposed on the extended H$\textsc{i}$ and thick disks of the nearby spiral galaxy M81.
Moreover, the low DM-excess of the FRB suggests that the FRB source is unlikely to be located beyond M81 group
($\sim$4 Mpc). We searched for galaxies closer than those associated with the M81 group within or near to the FRB localization region in the catalog of the local volume galaxies \citep{karachentsev2013updated} and found none. Additionally, the coincidence probability of finding an M81-like galaxy close to the FRB localization region is small ($< 1$\%). Therefore, if extragalactic, the FRB is most likely associated with M81 which would make it by far the closest extragalactic FRB yet known. Lastly, given the observational constraints, we cannot reject the Galactic origin of the FRB.

\subsection{Constraints on the Milky Way halo DM contribution}

Under the assumption that FRB 20200120E is extragalactic, it can be used to set an upper limit on the MW halo DM contribution in this direction.
If we consider the lowest of the two MW DM model estimates, DM$_{\mathrm{MW,YMW16}}$ = 35 pc cm$^{-3}$ and conservatively assume negligible IGM and host DM contribution,
then DM$_{\mathrm{MW,halo}} < 53$ pc cm$^{-3}$.  This would be inconsistent with most of the DM$_{\mathrm{MW,halo}}$ phase space proposed by \cite{prochaska2019low}, i.e., DM$_{\mathrm{halo}}$ = 50$-$80 pc cm$^{-3}$. On the other hand, both the \cite{dolag2015constraints} and \cite{yamasaki2020galactic} models predict DM$_{\mathrm{halo}} \sim~$30 pc cm$^{-3}$, lower than our upper limits. The halo may be clumpy \citep{Kaaret2020,keating2020exploring}, so it may still be possible to have significant variations in DM$_{\mathrm{halo}}$ along different sight-lines. More such low-DM FRB localizations will help in constraining the structure and composition of the MW halo.

\subsection{Comparison with SGR 1935+2154 radio bursts}

Table \ref{ta:bursts} provides the peak flux density of FRB 20200120E bursts. At a distance of 3.6 Mpc, the isotropic radio luminosity of the bursts would be $\sim~$10$^{37}$ erg~s$^{-1}$
similar to those of the very bright SGR 1935+2154 radio bursts recently detected by CHIME/FRB and STARE2 \citep{2020SGR,Bochenek2020}. 
In 2 yrs of CHIME/FRB observations,
we have seen at least three bursts from the FRB 20200120E source.
There are other low-DM FRBs within the CHIME/FRB sample that are presently under consideration.
Careful study of the host galaxy candidates in their error regions (which are presently mostly larger than for FRB 20200120E) must be done to assert them as extragalactic, given
the ever-present possibility of
them being in the distant MW or MW halo. This work is underway
and may be able to constrain the local volumetric FRB rate to compare with that inferred from SGR 1935+2154.   

The proximity of FRB 20200120E makes it an attractive target for X-ray and gamma-ray telescopes.  For a fiducial current
high-energy telescope fluence detection sensitivity threshold
of $10^{-10}$~erg~cm$^{-2}$, high-energy bursts from nearby sources with energies $> 10^{41}$~erg~s$^{-1}$ should be detectable, and are sometimes seen from Galactic magnetars in outburst \citep[see ][ for a review]{2017ARA&A..55..261K}.
Moreover, giant magnetar flares with total isotropic luminosity $\sim~$10$^{46}$ erg, like that from SGR 1806$-$20 \citep{palmer2005giant}, should be easily detectable by the {\it Swift}/Burst Alert Telescope (BAT) and {\it Fermi}/GBM which have 
flux
sensitivity of $\sim~$10$^{-7}$ erg cm$^{-2}$ s$^{-1}$ in the 15-150 keV band (A. Tohuvavohu, private
communication) and  $\sim~$10$^{-7}$ erg cm$^{-2}$ s$^{-1}$ in 50–300 keV band \citep{2020ApJ...893...46V}, respectively. Unfortunately,
{\it Fermi}/GBM was either not operational (transiting through the South Atlantic Anomaly region) or the FRB location was occulted by the Earth at all but one burst epoch. At the time of the 2020 February 6 burst, the FRB was visible to {\it Fermi}/GBM but no trigger was reported by the {\it Fermi} collaboration. Additionally, the {\it Swift}/BAT  field-of-view (FOV) did not cover the FRB sky-position at the time of FRBs 20200120E and 20201129A. 
However, the FRB location was within the {\it Swift}/BAT FOV at the time of FRB 20200718A, and no coincident X-ray burst was reported by the {\it Swift} collaboration. Therefore, if the FRB source is at 3.6 Mpc, it seems unlikely that FRB 20200718A and the 2020 February 6 event were associated with SGR~1806$-$20-like giant flares.


\subsection{Comparison with other repeating FRB hosts}

\begin{table}[ht]
\begin{center}
\caption{Notable properties of M81, the most promising host of FRB 20200120E.}
\label{m81-host-data}
\begin{tabular}{@{} *3l @{}}\toprule
\textbf{Property} & \textbf{Value} & \textbf{Reference}\\\midrule
SFR ($\textup{M}_\odot\ \mathrm{yr^{-1}}$) & 0.4$-$0.8 & \cite{gordon2004spatially} \\ 
Metallicity$^{a}$ [Z] (dex) & 0.03 &  \cite{2000AJ....119.2745K}\\
Stellar mass ($\rm{M}_{\odot}$) & (7.2 $\pm 1.7) \times 10^{10}$  & \cite{blokmetallicity2008}\\
Effective radius (R$_{\mathrm{eff}}$; kpc) &  3.5 & \cite{2010PASP..122.1397S}\\
(u-r)$_{0}$$^{b}$ (mag)& 2.773(4) & \cite{2009ApJS..182..543A} \\
E(V-B)$^{c}$ & 0.26 & \cite{kudritzki2012quantitative} \\
Absolute r-band mag. (AB) & $-$19.78 & -- \\
Inclination angle ($\degrees$) & 62   & \cite{karachentsev2013updated}\\
Luminosity distance (Mpc) & 3.63 $\pm$ 0.34 & \cite{karachentsev2013updated}\\
Projected FRB offset from galaxy center (kpc)$^{d}$ & 20$^{+3}_{-2}$  & This paper\\\bottomrule 
 \hline
\end{tabular}
\end{center}
$^{a}$ Average metallicity relative to the Sun i.e.,  
log(Z/Z$_{\odot}$). However, at the FRB location, the metallicity is found to be sub-solar, i.e., [Z] $<$ 0 \citep{2009AJ....137..419W}.\\
$^{b}$ Milky Way extinction is corrected using the reddening map by \cite{1998ApJ...500..525S}.

 $^{c}$ Average value of the color excess; at the FRB location, it is likely to be $< 0.1$ \citep{kudritzki2012quantitative}.
 
 $^{d}$ 90\% confidence interval.
\end{table}

In contrast with the late-type galaxy hosts of the only three other localized repeating FRBs,  FRBs 121102, 180916 and 190711 \citep{tendulkar2017host,marcote2020repeating,macquart2020census}, M81 is an early-type spiral galaxy of morphology SA(s)ab \citep{bosma198121}.
 Table \ref{m81-host-data} lists its main physical properties. 
 M81 also contains a low-luminosity AGN \citep{markoff2008results}, and is classified as a LINER Seyfert \citep{ho1996new}. \cite{heintz2020host} noted that the hosts of apparently non-repeating FRBs are typically more massive than those of repeating FRBs. However, M81 would be among the most massive FRB hosts known thus far$^{1}$
 with stellar mass of $7.2\times 10^{10} \;\textup{M}_\odot$ (see Table \ref{m81-host-data}). 
 Lastly, if we ignore FRB 190523 for which the host association is not firm \citep{macquart2020census,heintz2020host}, FRB 20200120E would show the largest projected offset from the center of its host ($\sim$~20 kpc). This would be at odds with the offset distribution of the progenitors of long gamma-ray bursts and superluminous supernovae, which are found close to the centers of their respective hosts \citep{lunnan2015zooming,blanchard2016offset,heintz2020host,Mannings+21}.
 
 If FRB 20200120E is a classical magnetar, it would be surprising to find it at such a large offset from the center of its host; known Galactic magnetars are all well within the optical disk of the  Milky Way \citep{olausen2014mcgill}.
 This same issue has been noted for other localized FRBs \citep{heintz2020host}.
 However, 
 the M81 circumgalactic medium (CGM) is dynamic and rich in gas and metals \citep{chen2017outskirts}. 
 \cite{sun2005intergalactic} and \cite{2019arXiv191014672S} noted the existence of a diffuse stellar population embedded in the extended H$\textsc{i}$ disk where in situ star formation is actively taking place. Lastly, \cite{frederiks2007possibility} and \cite{2010MNRAS.403..342H} have argued for the existence of a neutron star population, including a possible magnetar, 
 in the CGM of M81. 
  
\section{Conclusions}
\label{Sec:conclude}
We have reported on the detection of the repeating fast radio burst source FRB 20200120E discovered with CHIME/FRB.  This source has very low DM, 87.82 pc cm$^{-3}$, though greater than what is expected from models of the Milky Way ISM along its line-of-sight. Due to large uncertainties in the Milky Way halo DM contribution, it is possible that the FRB source is within our halo. However, we find no cataloged Milky Way halo satellite galaxy or globular cluster within or near to the FRB localization region that can host the FRB source. Moreover, the presence of a solitary neutron star capable of producing FRB-like radio emission at a distance $\sim~50-200$~kpc seems unlikely. On the other hand, we identify M81, a nearby grand-design spiral galaxy at a distance of $\sim$~3.6 Mpc, with an angular offset $\approx 19\arcmin$ 
 and chance coincidence probability $< 10^{-2}$, making it a promising host candidate.

We have shown that the observed extragalactic DM component of the FRB is significantly lower than the model-predicted DM contribution from the M81 halo as a foreground galaxy.
This suggests that the FRB host galaxy is unlikely to be located beyond M81, though the FRB may exist within the extended disk of M81.
Therefore, if extragalactic, FRB 20200120E is most likely associated with M81.
M81 is different from the hosts of other known repeating FRBs in spatial offset, stellar population age, and local environment.  This supports an interesting diversity in repeater host properties that additional localizations will help understand.  
We also found that the FRB localization region contains the extended M81 HI-disk and a number of interesting M81 sources including an H\textsc{ ii} region ([PWK2012] 31), an X-ray binary ([SPZ2011] 8) and a VLASS source, VLASS1QLCIR J095756.10+684833.3.
At a distance of 3.6 Mpc, it should be possible to detect prompt multi-wavelength counterparts of FRB 20200120E predicted by several FRB models, including the magnetar model. Some FRB models anticipate even greater luminosities in high energy bands than in the radio band \citep{yi2014multi,burke2018multiple,chen2020multiwavelength}. For example, in the synchrotron maser model of \cite{metzger2019fast}, shock-heated electrons gyrate to produce synchrotron radiation that sweeps through the $\gamma$-ray and X-ray bands, and in some cases, even extends to the optical band on sub-second timescales. Additionally, if radio bursts from
FRB 20200120E are accompanied by X-ray bursts, as was seen in SGR 1935+2154 \citep{2020SGR,Bochenek2020,mereghetti2020integral,li2020identification,ridnaia2020peculiar}, detecting a coincident X-ray counterpart seems feasible, and
would be a strong test of the magnetar origin of FRBs. Therefore, we encourage multi-wavelength follow-up of FRB 20200120E.

\acknowledgements

We thank the anonymous reviewer and ApJL statistics editor for their careful reading of our manuscript and their many insightful comments and suggestions. We expression our gratitude to M. Rahman and J. X. Prochaska  for the useful discussions.
We are grateful to the staff of the Dominion Radio Astrophysical Observatory, which is operated by the National Research Council Canada.
The CHIME/FRB Project is funded by a grant from the Canada Foundation for Innovation 2015 Innovation Fund (Project 33213), as well as by the Provinces of British Columbia and Quebec, and by the Dunlap Institute for Astronomy and Astrophysics at the University of Toronto.
Additional support was provided by the Canadian Institute for Advanced Research (CIFAR) Gravity \& Extreme Universe Program, McGill University and the McGill Space Institute and the University of British Columbia. The Digitized Sky Surveys were produced at the Space Telescope Science Institute under U.S. Government grant NAG W-2166. The images of these surveys are based on photographic data obtained using the Oschin Schmidt Telescope on Palomar Mountain and the UK Schmidt Telescope. The plates were processed into the present compressed digital form with the permission of these institutions.
The National Geographic Society - Palomar Observatory Sky Atlas (POSS-I) was made by the California Institute of Technology with grants from the National Geographic Society.
The Second Palomar Observatory Sky Survey (POSS-II) was made by the California Institute of Technology with funds from the National Science Foundation, the National Geographic Society, the Sloan Foundation, the Samuel Oschin Foundation, and the Eastman Kodak Corporation.
The Oschin Schmidt Telescope is operated by the California Institute of Technology and Palomar Observatory.
The UK Schmidt Telescope was operated by the Royal Observatory Edinburgh, with funding from the UK Science and Engineering Research Council (later the UK Particle Physics and Astronomy Research Council), until 1988 June, and thereafter by the Anglo-Australian Observatory. The blue plates of the southern Sky Atlas and its Equatorial Extension (together known as the SERC-J), as well as the Equatorial Red (ER), and the Second Epoch [red] Survey (SES) were all taken with the UK Schmidt.
All data are subject to the copyright given in the copyright summary. Copyright information specific to individual plates is provided in the downloaded FITS headers. Supplemental funding for sky-survey work at the ST ScI is provided by the European Southern Observatory. M.B. is supported by an FRQNT Doctoral Research Award. The Dunlap Institute is funded through an endowment established by the David Dunlap family and the University of Toronto. B.M.G. acknowledges the support of the Natural Sciences and Engineering Research Council of Canada (NSERC) through grant RGPIN-2015-05948, and of the Canada Research Chairs program. V.M.K. holds the Lorne Trottier Chair in Astrophysics \& Cosmology, a Distinguished James McGill Professorship and receives support from an NSERC Discovery Grant (RGPIN 228738-13) and Gerhard Herzberg Award, from an R. Howard Webster Foundation Fellowship from CIFAR, and from the FRQNT CRAQ. D.M. is a Banting Fellow. P.C. is supported by an FRQNT Doctoral Research Award.
C.L. was supported by the U.S. Department of Defense (DoD) through the National Defense Science \& Engineering Graduate Fellowship (NDSEG) Program.
M.D. receives support from a Killam fellowship, NSERC Discovery Grant,
CIFAR, and from the FRQNT Centre de Recherche en
Astrophysique du Quebec.
P.S. is a Dunlap Fellow and an NSERC Postdoctoral Fellow. 
K.S. acknowledges support by the NSF Graduate Research Fellowship Program.
The baseband system is funded in part by a CFI JELF award to I.H.S.


\bibliographystyle{aasjournal}
\bibliography{ref.bib}
\end{document}